\documentclass[twocolumn,british,aps,superscriptaddress,prc,nofootinbib,showpacs,showkeys]{revtex4-1}
\usepackage[T1]{fontenc}
\usepackage[latin9]{inputenc}
\usepackage{color}
\usepackage{babel}
\usepackage{amsmath}
\usepackage{graphicx}
\usepackage{esint}
\usepackage[unicode=true,pdfusetitle,
 bookmarks=true,bookmarksnumbered=false,bookmarksopen=false,
 breaklinks=false,pdfborder={0 0 1},backref=false,colorlinks=true]
 {hyperref}
\hypersetup{
 pdfborderstyle=,linkcolor=blue,citecolor=cyan}

\makeatletter

\providecommand{\tabularnewline}{\\}

\usepackage{babel}

\usepackage{xcolor}

\makeatother

\begin{document}
\title{The effect of electric and chiral magnetic conductivities on azimuthally
fluctuating electromagnetic fields and observables in isobar collisions}
\author{Irfan Siddique}
\email{irfansiddique@ucas.ac.cn.edu.cn}

\affiliation{School of Nuclear Science and Technology, University of Chinese Academy
of Sciences, Beijing 101408, China}
\author{Uzma Tabassam}
\affiliation{Department of Physics, COMSATS University Islamabad Campus, Islamabad,
Park Road, 44000 Pakistan}
\begin{abstract}
We study the space-time evolution of electromagnetic fields along
with the azimuthal fluctuations of these fields and their correlation
with the initial matter geometry specified by the participant plane
in the presence of finite electric $\left(\sigma\right)$ and chiral
magnetic $\left(\sigma_{\chi}\right)$ conductivities in Ru+Ru and
Zr+Zr collisions at $\sqrt{s_{NN}}=200$ GeV. We observe the partially
asymmetric behavior of the spatial distributions of the electric and
magnetic fields in a conducting medium when compared to the Lienard-Wiechert
(L-W) solutions, and deceleration of the decay of the fields is observed
in both isobar collisions. While studying the correlation between
the magnetic field direction and the participant plane, we see the
sizeable suppression of the correlation in the presence of finite
conductivities when compared to the L-W case, reflecting the importance
of taking into account the medium properties such as conductivities
while calculating the magnetic field induced observable quantities. 
\end{abstract}
\maketitle

\section{Introduction}

The Ultra-relativistic heavy ion collision creates a deconfined state
with extremely high energy and density known as Quark-Gluon-Plasma
(QGP). In non-central heavy ion collisions, along with this high energy
and density state, very strong electromagnetic fields are also generated
due to charged particles having relativistic motions, which provides opportunity
 to study related phenomenon in heavy ion collisions. Typical
strength of magnetic field produced in heavy ion collisions can be
estimated in co-moving frame for fast moving nuclei by $eB\sim\gamma Ze^{2}/R_{A}^{2}$,
with $\gamma$ being the Lorentz Factor, $Z$ being the proton number
and $R_{A}$ being the radius of the nucleus. For the case of Au+Au
collisions at top RHIC (Relativistic Heavy ion collider) energies
$(\sqrt{s_{NN}}=200\text{ GeV})$ magnetic field is of orders of $eB\sim m_{\pi}^{2}\sim10^{18}$
Gauss \citep{Bzdak:2011yy,Deng:2012pc,Voronyuk2011} and fields are
proportional to collision energy so at LHC (Large Hadron Collider)
energies in Pb+Pb collisions at $(\sqrt{s_{NN}}=2.76\ \text{TeV})$
it can be roughly 10 times more stronger \citep{Skokov2009,Zhong:2014cda}.
In recent years, many developments and efforts have been made to explore
the effects induced by (electro)magnetic fields such as the Chiral
Magnetic Effect (CME) \citep{Kharzeev:2007jp,Kharzeev:2007tn,Fukushima:2008xe,Fukushima:2010vw,Kharzeev2011,Son:2009tf},
the Chiral Separation Effects (CSE) \citep{Son:2004tq,Metlitski:2005pr},
Chiral Magnetic Wave (CMW) \citep{Kharzeev:2010gd,Burnier:2011bf,Burnier:2012ae,Yee:2013cya,Jiang2015}
etc. All of these effects are related to chiral fermions or massless
fermions. Search for the CME is currently a very active field of interest
in heavy ion collisions at the RHIC and LHC \citep{Abelev:2009uh,Abelev:2009txa,Abelev2013,Adamczyk2013,Adamczyk2014,Adamczyk2014a,Khachatryan2017}.
In early STAR and ALICE experiments charge separation effect was measured
by measuring two particle azimuthal angle correlation $\gamma_{\alpha\beta}=\left\langle \cos\left(\phi_{\alpha}+\phi_{\beta}-2\Psi_{RP}\right)\right\rangle $
with $\phi_{i}$ being the azimuthal angle of corresponding charged
particle, $\alpha$ $\left(\beta\right)$ denotes the sign of charge
particle (either positive or negative) and $\Psi_{RP}$ being the
reaction plane angle, and the measurements support the presence of
CME \citep{Abelev:2009uh,Abelev:2009txa,Abelev2013}. But due to inseparable
contribution from the background, it is extremely difficult to properly
understand and extract the CME signal from the huge background in
experiment results \citep{Schlichting2011,Wang2017,Zhao2019,Adamczyk2014,Khachatryan2017}.
There have been several attempts to eliminate or reduce background
effects \citep{Bzdak2013,Zhao:2017nfq,Xu2018,Wang2010}.

According to the expectations from CME, the difference between the
correlation of opposite charge pairs $\left(\gamma_{oppsite}\right)$
and same charge pairs $\left(\gamma_{same}\right)$ is expected to
be directly proportional to the strength of the squared magnetic field
$\left(e\text{\textbf{B}}\right)^{2}$ and azimuthal fluctuations
of the magnetic field direction $\left(\cos2\left(\Psi_{B}-\Psi_{2}\right)\right)$
\citep{Bloczynski:2012en,Bloczynski2015,Chatterjee2015} i.e, 
\begin{equation}
\Delta\gamma=\gamma_{opposite}-\gamma_{same}\propto\left\langle \left(e\text{\textbf{B}}\right)^{2}\cos2\left(\Psi_{B}-\Psi_{2}\right)\right\rangle \label{eq:del-gama}
\end{equation}
where $\Psi_{B}$ represents the azimuthal angle of the magnetic field
and $\Psi_{2}$ represents the second harmonic participant plane angle.
The right hand side of the above equation shows that the quantitative
contribution of the B-field-induced effect is essentially controlled
by $\left(e\text{\textbf{B}}\right)^{2}\cos2\left(\Psi_{B}-\Psi_{2}\right)$,
therefore this projected field strength controls the contribution
rather than $\left(e\text{\textbf{B}}\right)^{2}$ alone.

To extract the CME signal, a solution proposed is to carry out isobaric
collisions $\text{Ru}_{44}^{96}+\text{Ru}_{44}^{96}$ ($\text{Ru: }$Ruthenium)
and $\text{Zr}_{40}^{96}+\text{Zr}_{40}^{96}$ ($\text{Zr: }$Zirconium)
\citep{Voloshin2010} and experiments were performed in RHIC \citep{STAR:2021mii,Hu2022,Hu2023}
along this line to observe these effects. These isobaric collisions
are intensely pursued for investigation because the advantage is that
the difference in number of protons can generate different magnitudes
of electromagnetic fields and related induced effects, but the same
mass number in two isobar systems can generate the same background
effect. So one can expect to observe the CME signal if it really exists
in heavy ion collisions. For instance, from Woods-Saxon distributions
it can be confirmed that in isobaric collisions $e\boldsymbol{\textbf{B}}$
differs by $10\%$ \citep{Deng2016}, which naively agrees to the
fact that the atomic number in Ru and Zr differ by 10\% (Ru-44, Zr-40),
so there can be a chance of observing a CME signal according to Eq.
\ref{eq:del-gama}. There are several studies that show framework
to search for CME and predict for the correlation observables by using
the initial magnetic field produced in isobar collisions i.e., introducing
the initial charge separation proportional to magnetic field in AMPT
model and studying the effect of final state interactions on CME observables
\citep{Deng2018}, detecting CME signal and predicting for the correlation
observables by using absolute difference between two isobars event
with identical multiplicity and elliptic-flow in Anomalous-Viscous
Fluid Dynamics (AVFD) framework \citep{Shi2019}, measurement of $\Delta\gamma$
with respect to reaction plane $\left(\Psi_{RP}\right)$ and participant
plane $\left(\Psi_{pp}\right)$ and compare between them by using
MC glauber and AMPT model \citep{Xu2018}, reflecting information
about CME by studying the correlation between (initial) magnetic field
direction and second harmonics of participant plane angle $\Psi_{pp}$
(and spectator plane $\Psi_{sp}$ ) \citep{Zhao2019a}. In previous
studies, magnetic field without medium feedback is used for predicting
about CME and correlation observables however in this study we follow
\citep{Zhao2019a} and perform analysis for a more realistic nuclear
matter that take into account medium feedback in terms of electric
and chiral magnetic conductivities. 

Supposing that two isobar systems have same background, then $\Delta\gamma$
is expecting main contributions from the squared magnetic field and
the correlation between the azimuthal angle of the magnetic field
$\Psi_{B}$ and participant plane $\Psi_{2}$. So the first key ingredient
is the fact that magnetic field whose spatial and time evolution in
different mediums can behave differently \citep{McLerran:2013hla,Tuchin:2013apa,Tuchin:2014iua,Li:2016tel,Siddique:2021smf,Siddique2022},
so consequently can have effect on related observables. The chiral
conductivity being directly proportional to the chiral chemical potential
fluctuates in each event. When computing event-averaged observables,
the intrinsic diversity in the chiral chemical potential for each
event presents a difficulty in directly determining its impact on
the observable. However, in the context of event-by-event analysis,
each unique event contains the distinct influence of the chiral conductivity
on the observable. In contrast to the case where fluctuations are
averaged out over several events, the event-by-event analysis enables
us to capture and study the subtle influence of variations in the
chiral chemical potential on the observable. Although it is difficult
to see the direct influence of chiral chemical potential while calculating
event-averaged observable but in event-by-event analysis each event
encodes its effect in calculating observable so its effect is not
totally smoothed out. So in this paper, we will look at the electric
and magnetic fields produced in isobar collisions in the presence
of electric and chiral magnetic conductivities and measure their effect
on the azimuthal fluctuations of electromagnetic fields and related
observable quantities.

After providing the brief introduction, in Section \ref{sec:2} we
give the expressions for electric and magnetic fields in zero conductivity
system and system having finite conductivity. In Section \ref{sec:3}
we provide simulation results and discussions for the electromagnetic
fields and their correlation with initial matter geometry specified
by the participant plane in central and non central isobaric collisions.
Finally we summarize in Section \ref{sec:summary}.

\section{\label{sec:2} Calculation of electromagnetic field}

\subsection*{A: Zero-conductivity system ($\sigma=\sigma_{\chi}=0$)}

For a system with zero conductivity, or vacuum ($\sigma=\sigma_{\chi}=0$),
the electric and magnetic field in each event can be evaluated by
using the Lienard-Wiechert potential~\citep{Bzdak:2011yy,Bloczynski:2012en}
as 
\begin{equation}
{\textbf{E}}(t,\textbf{x})=\alpha_{EM}\sum_{n}\frac{\left(1-v_{n}^{2}\right)\textbf{R}_{n}}{\left(\textbf{R}_{n}^{2}-\left(\textbf{R}_{n}\times\textbf{v}_{n}\right)^{2}\right)^{3/2}},\label{eq:LW_E}
\end{equation}
\begin{equation}
{\textbf{B}}(t,\textbf{x})=\alpha_{EM}\sum_{n}\frac{\left(1-v_{n}^{2}\right)\left(\textbf{v}_{n}\times\textbf{R}_{n}\right)}{\left(\textbf{R}_{n}^{2}-\left(\textbf{R}_{n}\times\textbf{v}_{n}\right)^{2}\right)^{3/2}},\label{eq:LW_B}
\end{equation}
where $\textbf{R}_{n}=\textbf{x}-\textbf{x}_{n}$ is the relative
position vector between the source point $\textbf{x}_{n}$ and the
field point $\textbf{x}$ under discussion, and $\textbf{x}_{n}$
and $\textbf{v}_{n}$ represent the position and velocity respectively
of the $n$-th proton in the colliding nuclei at the current time
$t$. In above equation $\alpha_{EM}=e/4\pi\approx1/137$ is fine
structure constant. Note that the Eqs. (\ref{eq:LW_E}) and (\ref{eq:LW_B})
are valid under the assumption that all the source charges are traveling
with a constant velocity. If all the charges do not have constant
velocity then the original form of the Lienard-Wiechert fields \citep{Deng:2012pc,Siddique:2021smf}
using the retarded time should be used for the calculation of the
electric and magnetic fields.

\subsection*{B: Conducting system with finite conductivities ($\sigma\protect\neq0,\sigma_{\chi}\protect\neq0$)}

The quark-gluon plasma (QGP) matter is produced in heavy-ion collisions
and it has certain conducting property so it is important to take
into account the feedback effects of the conductivities. The Maxwell
equations with both electric ($\sigma$) and chiral magnetic ($\sigma_{\chi}$)
conductivities can be written in the following form: 
\begin{eqnarray}
\nabla\cdot\textbf{F} & = & \left\{ \begin{array}{cc}
\rho_{\mathrm{ext}}/\epsilon & \rightarrow\textbf{E}\\
0 & \rightarrow\textbf{B}
\end{array}\right.,\label{eq:eq1}\\
\nabla\times\textbf{F} & = & \left\{ \begin{array}{cc}
-\partial_{t}\textbf{B}\text{ \ \ \ \ \ \ } & \rightarrow\textbf{E}\\
\partial_{t}\textbf{E}+\textbf{J}_{\mathrm{3}}(\sigma,\sigma_{\chi})\ \ \  & \rightarrow\textbf{B}
\end{array}\right.,\label{eq:eq1.2}
\end{eqnarray}
where $\rho_{\mathrm{ext}}$ is external charge density and $\textbf{J}_{\mathrm{3}}(\sigma,\sigma_{\chi})=\textbf{J}_{\mathrm{ext}}+\sigma\textbf{E}+\sigma_{\chi}\textbf{B}$
with $\textbf{J}_{\mathrm{ext}}$ being external current density,
$\sigma\textbf{E}$ being electric current and $\sigma_{\chi}\textbf{B}$
being chiral current, and $\textbf{F}$ denotes either electric ($\textbf{E}$)
or magnetic ($\textbf{B}$) field. We can obtain the following algebraic
expressions for electric and magnetic field components with finite
electric and chiral conductivity by solving the above Maxwell equation
using the Green's function method in cylindrical coordinates by considering
that all source charges fly along the $z$-axis~\citep{Li:2016tel}:
\begin{align}
B_{\phi}(t,\textbf{x}) & =\frac{Q}{4\pi}\frac{v\gamma x_{\mathrm{T}}}{\Delta^{3/2}}\left(1+\frac{\sigma v\gamma}{2}\sqrt{\Delta}\right)e^{A},\nonumber \\
B_{r}(t,\textbf{x}) & =-\sigma_{\chi}\frac{Q}{8\pi}\frac{v\gamma^{2}x_{\mathrm{T}}}{\Delta^{3/2}}e^{A}\left[\gamma\left(vt-z\right)+A\sqrt{\Delta}\right],\nonumber \\
B_{z}(t,\textbf{x}) & =\sigma_{\chi}\frac{Q}{8\pi}\frac{v\gamma}{\Delta^{3/2}}e^{A}\Big[\Delta\left(1-\frac{\sigma v\gamma}{2}\sqrt{\Delta}\right)\nonumber \\
 & \;\;\;\;+\gamma^{2}\left(vt-z\right)^{2}\left(1+\frac{\sigma v\gamma}{2}\sqrt{\Delta}\right)\Big],\label{eq:Brphiz}
\end{align}
in which $\Delta$ is defined as $\Delta\equiv\gamma^{2}\left(vt-z\right)^{2}+x_{\mathrm{T}}^{2}$
and $A$ is defined as $A\equiv\left(\sigma v\gamma/2\right)\left[\gamma\left(vt-z\right)-\sqrt{\Delta}\right]$;
and 
\begin{align}
E_{\phi}(t,\textbf{x}) & =\sigma_{\chi}\frac{Q}{8\pi}\frac{v^{2}\gamma^{2}x_{\mathrm{T}}}{\Delta^{3/2}}e^{A}\left[\gamma\left(vt-z\right)+A\sqrt{\Delta}\right],\nonumber \\
E_{r}(t,\textbf{x}) & =\frac{Q}{4\pi}e^{A}\Bigg\{\frac{\gamma x_{\mathrm{T}}}{\Delta^{3/2}}\left(1+\frac{\sigma v\gamma}{2}\sqrt{\Delta}\right)\nonumber \\
 & \;\;\;\;-\frac{\sigma}{vx_{\mathrm{T}}}e^{-\sigma(t-z/v)}\left[1+\frac{\gamma\left(vt-z\right)}{\sqrt{\Delta}}\right]\Bigg\},\nonumber \\
E_{z}(t,\textbf{x}) & =\frac{Q}{4\pi}\left\{ -\frac{e^{A}}{\Delta^{3/2}}\left[\gamma\left(vt-z\right)+A\sqrt{\Delta}+\frac{\sigma\gamma}{v}\Delta\right]\right.\nonumber \\
 & \;\;\;\;\left.+\frac{\sigma^{2}}{v^{2}}e^{-\sigma\left(t-z/v\right)}\Gamma\left(0,-A\right)\right\} ,\label{eq:Erphiz}
\end{align}
where $\Gamma\left(0,-A\right)$ is the incomplete gamma function
defined as $\Gamma\left(a,z\right)=\int_{z}^{\infty}t^{a-1}\exp\left(-t\right)\,dt$.
It can be verified that with zero conductivity limit, Eqs.~(\ref{eq:Brphiz})
and~(\ref{eq:Erphiz}) return to the Lienard-Wiechert solution.

We have used the Monte-Carlo (MC) Glauber model developed by the PHOBOS
collaboration \citep{Loizides:2014vua} to calculate the spatial distribution
of the source nucleons. MC Glauber model are useful for estimating
EM fields in heavy ion collisions, and also with the help of Glauber
model computations we can connect experimental data to collision centrality
and other geometric quantities such as eccentricity, initial state
anisotropy etc \citep{Alver:2006wh,Miller:2007ri,Alver:2008zza,Siddique2022}.
To determine the spatial information of the source charges, a two-step
calculation is performed in this model. The reaction plane is defined
by the impact parameter and the beam direction, represented by $x$-axis
and $z$-axis respectively. First, the centers of projectile and target
nuclei are located at $x=\pm b/2$ for a given impact parameter $b$,
and then it is employed to determine spatial positions of nucleons
stochastically in the two colliding nuclei. The density profile of
isobar nucleus have following Woods-Saxon (WS) distribution 
\begin{equation}
\rho(x,y,z)=\frac{\rho_{0}}{1+\exp\left[\frac{r-R(1+\beta_{2}Y_{20}+\beta_{4}Y_{40})}{f}\right]},
\end{equation}
where $\rho_{0}$ is the nuclear density at the nucleus center, $f$
represents the surface thickness parameter, and $Y_{nl}(\theta)$
represents spherical harmonic functions. Here $\beta_{2}$, and $\beta_{4}$
are deformation parameters and they determine the deviation from a
spherical shape of the nucleus. Taking into account these parameters
allows us to obtain more realistic representations of spatial distribution
of nucleons within the projectile and target nuclei. After determining
the Woods-Saxon distribution information, the subsequent motions are
then assumed to be along straight trajectories in the beam direction
(i.e., in $+$/$-z$ directions). The collision between nucleons from
projectile and target occurs if they satisfy $d\leq\sqrt{\sigma_{\mathrm{inel}}^{\mathrm{NN}}/\pi}$,
where $d$ is the distance in the transverse plane and $\sigma_{\mathrm{inel}}^{\mathrm{NN}}$
denotes the inelastic cross section of nucleon-nucleon collisions.
Those nucleons that do not partake in collisions are labeled as ``spectators''
while those that partake in collisions are labeled as ``participants''.

\begin{figure}
\includegraphics[width=8cm]{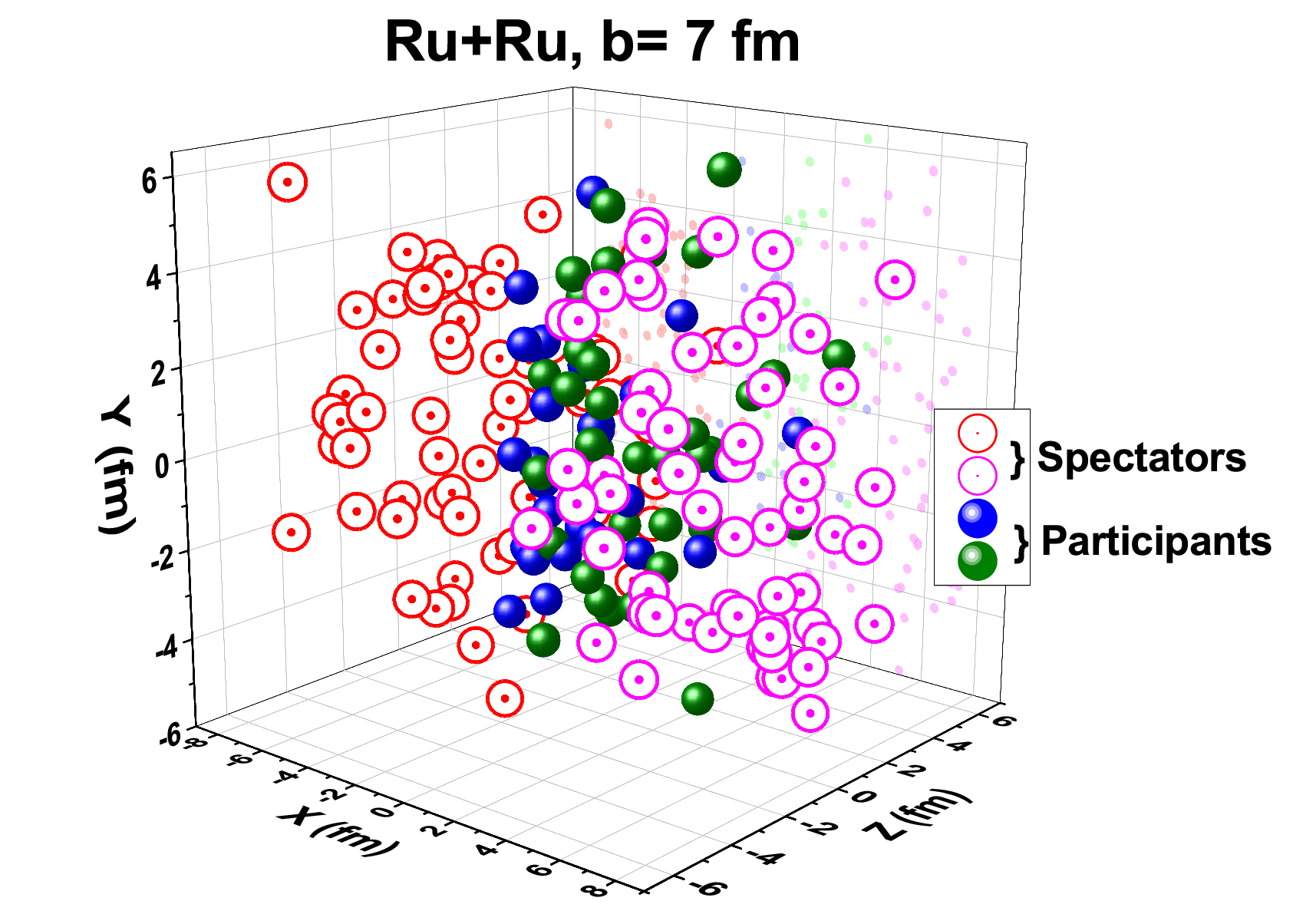}

\caption{\label{fig:Initial-geometry}Initial geometry of a Ru+Ru collision
event generated by the Monte-Carlo Glauber model for $b=7$~fm and
$\sqrt{s_{\mathrm{NN}}}=200$~GeV on the 3D plane.}
\end{figure}

In this work, we have used two isobar systems i.e., $_{44}^{96}\text{Ru\ensuremath{+_{44}^{96}}Ru}$
(Ruthenium-96) and $_{40}^{96}\text{Zr\ensuremath{+_{40}^{96}}Zr}$
(Zirconium-96) collisions. There are several WS parameter settings
for two isobar nuclei. For the case of deformed nuclei structure,
we use $R=5.085$~fm and 5.020 fm for Ru and Zr respectively, and
$d=0.46$~fm for both Isobar collisions at $\sqrt{s_{\mathrm{NN}}}=200$~GeV
at RHIC. The deformation parameter $\beta_{2}$ is not confirmed yet
and there are two cases for $\beta_{2}$ \citep{Moller1995,Pritychenko2016,Shou2015}.
In one situation deformation parameter for $_{44}^{96}\text{Ru}$
is larger then $_{40}^{96}\text{Zr}$ ($\beta_{2}^{Ru}=0.158$ and
$\beta_{2}^{Zr}=0.08$), while in other deformation parameter for
$_{44}^{96}\text{Ru}$ is smaller than $_{40}^{96}\text{Zr}$ ($\beta_{2}^{Ru}=0.053$
and $\beta_{2}^{Zr}=0.217$) but in this paper we have only considered
the first situation because the effect of deformation parameter has
been studies in Ref. \citep{Zhao2019a}. Apart from deformed nuclei
structure there are also other WS distribution which can reflect neutron-skin
effect, details of these parameters settings and effects are given
and studied in \citep{Xu2021,Zhao2022}. Among several WS parameter
settings studied in \citep{Zhao2022} halotype nuclei has reproduce
best experimental results for average number of charge particles,
charged particle multiplicity ($N_{\text{ch}}$) and elliptical flow
so we also considered halotype nuclei structure alongwith deformed
nuclei structure. WS parameters used in this study are given in table
\ref{tab:Woods-Saxon-parameters-for}.

\begin{table}
\begin{tabular}{|c|c|c|c|}
\hline 
\multicolumn{4}{|c|}{Deformed Nuclei Case \citep{Moller1995,Pritychenko2016,Shou2015}}\tabularnewline
\hline 
\hline 
 & $R_{0}$ & $a$ & $\beta_{2}$\tabularnewline
\hline 
Ru & 5.085 & 0.46 & 0.158\tabularnewline
\hline 
Zr & 5.020 & 0.46 & 0.08\tabularnewline
\hline 
\multicolumn{4}{|c|}{Halotype Nuclei Case \citep{Xu2021,Zhao2022}}\tabularnewline
\hline 
Ru, n & 5.085 & 0.523 & 0\tabularnewline
\hline 
Ru, p & 5.085 & 0.523 & 0\tabularnewline
\hline 
Zr, n & 5.021 & 0.592 & 0\tabularnewline
\hline 
Zr, p & 5.021 & 0.523 & 0\tabularnewline
\hline 
\end{tabular}

\caption{\label{tab:Woods-Saxon-parameters-for}Woods-Saxon parameters for
Ru and Zr}
\end{table}

In Fig.~\ref{fig:Initial-geometry}, we show our initial charge distribution
based on the Phobos MC Glauber on the 3D plane together with their
XZ projections on the XY plane for $_{40}^{96}\text{Ru-}{}_{40}^{96}\text{Ru}$
with $b=7\text{ fm}$. Solid blue and green circles represent participant
nucleons from the two colliding nuclei, while red and magenta open
circles represent spectators that do not participate in inelastic
scatterings. In order to assess whether a nucleon (proton) contributes
to the electromagnetic field, we utilize the probability $Z/A$ ($44/96$
for the Ru nucleus and 40/96 for the Zr nucleus). The evaluation of
the electromagnetic field takes into account the contributions from
the protons in both participants and spectators. It has been found
that there is a tiny difference in magnitudes between considering
only spectators and all nucleons. For instance, in Ru+Ru collisions
at $b=9\text{ fm}$, protons in the spectator alone produce around
4\% less $B_{y}$ than protons in the entire nucleus and almost similar
yield is observed as well for Zr+Zr collisions in our setup. For calculating
the electromagnetic field in the rest of this study, we perform event-by-event
analysis over 100,000 MC Glauber events for each impact parameter
(number of participants) setup to obtain a event-by-event geometric
distribution of the source charges and relevant results.

The main objective is to study the correlation between electromagnetic
fields and plane angle formed by participants for zero-conductivity
and finite conductivity cases, we use the definition of the $n\text{th}$
plane participant plane angle represented by $\Psi_{n}$ which can
be calculated as

\begin{equation}
\Psi_{n}=\frac{\text{atan}2\left(\left\langle r_{P}^{2}\text{sin}\left(n\phi_{P}\right)\right\rangle ,\left\langle r_{P}^{2}\text{cos}\left(n\phi_{P}\right)\right\rangle +\pi\right)}{n},\label{eq:psin}
\end{equation}
where $n$ can be 1,2,3 ... representing different flow harmonics,
however in this work we only consider $n=2$ because the second harmonics
are most prominent and are tightly correlated to initial geometry
distributions, $r_{P}$ represents the displacement of participating
nucleons from the field point $\boldsymbol{r}$ and $\phi_{P}$ representing
their corresponding azimuthal angle on the transverse plane. We also
check the similarity and dissimilarity between zero-conductivity limit
and finite conductivity system by calculating relative ratios defined
by 
\begin{equation}
X_{c}=2\frac{c^{Ru}-c^{Zr}}{c^{Ru}+c^{Zr}}
\end{equation}
where $c$ represents correlation $\left\langle \cos2\left(\Psi_{\text{F}}-\Psi_{2}\right)\right\rangle $
or $\left\langle e\text{\textbf{F}}^{2}\cos2\left(\Psi_{\text{F}}-\Psi_{2}\right)\right\rangle $
in our calculations. $X_{c}$ close to zero represents similarity
while away from zero represents dissimilarity between two isobar collision
systems. Also by studying relative ratio trends we can get information
about qualitative difference in zero-conducting case and finite conductivities
case. If trends differ from each other more it will show difference
in qualitative behavior in vacuum scenario and finite conductivity
scenario.

\section{Simulation results and Discussions\label{sec:3}}

\subsection{Effects on Electromagnetic fields}

\subsubsection{Spatial Distributions:}

In this subsection we show the numerical results of the contour plots
of electric and magnetic fields compared between zero-conducting system
and finite conductivity system (which have finite values of electric
and chiral magnetic conductivity) for isobar collisions. As described
in the previous section, the space-time evolution profile for electric
charges for Ru+Ru and Zr+Zr is used for collisions at $\sqrt{s_{NN}}=200$
GeV for different impact parameters. We use the Lienard-Wiechert solution
for the case of zero-conductivity $\left(\sigma=\sigma_{\chi}=0\right)$,
while for finite conductivity case Maxwell equations are solved to
obtain the electromagnetic fields.

\begin{figure*}
\includegraphics[width=12cm]{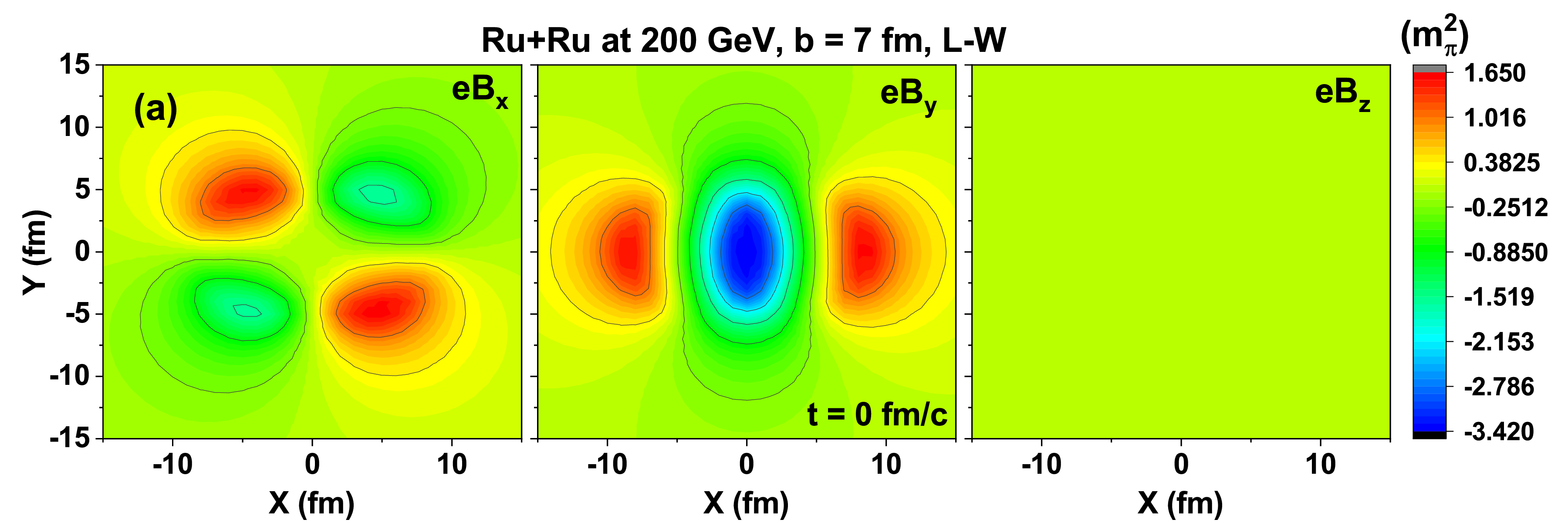}

\includegraphics[width=12cm]{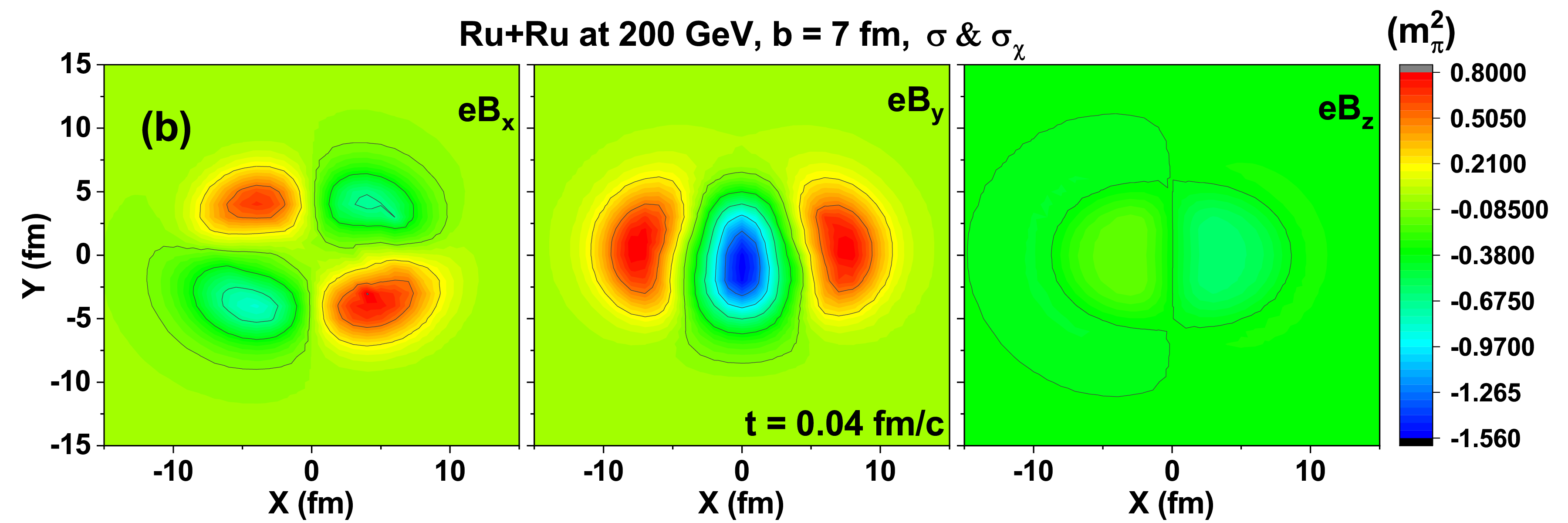}

\caption{\label{fig:spatial-B}Spatial distribution of a magnetic field components
($eB_{x,y,z}$ in unit of $m_{\pi}^{2}$) in Ru+Ru for $b=7$ fm at
$\sqrt{s_{NN}}=200$ GeV, compared between zero conductivity case
(first row) vs finite conductivities case (second row).}
\end{figure*}

\begin{figure*}
\includegraphics[width=12cm]{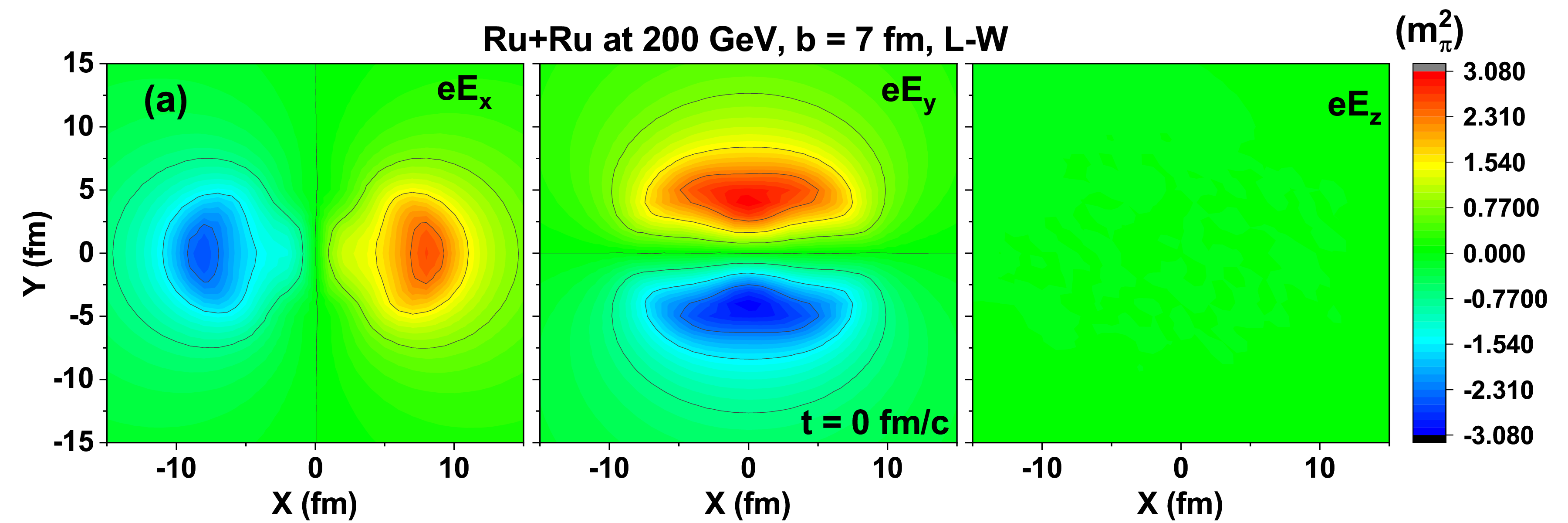}

\includegraphics[width=12cm]{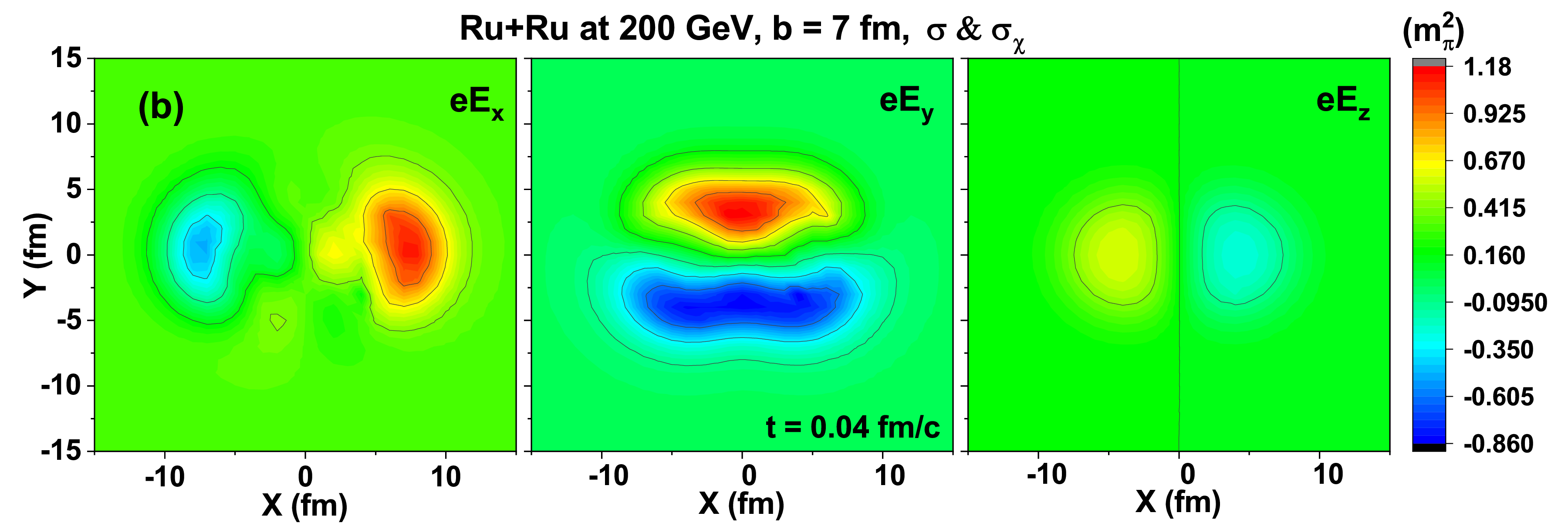}

\caption{\label{fig:spatial-E}Spatial distribution of a electric field components
( $eE_{x,y,z}$ in unit of $m_{\pi}^{2}$) in Ru+Ru for $b=7$ fm
at $\sqrt{s_{NN}}=200$ GeV, compared between zero conductivity case
(first row) vs finite conductivities case (second row).}
\end{figure*}

In Fig. \ref{fig:spatial-B} and \ref{fig:spatial-E}, we show the
contour plots of magnetic field $\left\langle B_{x,y,z}\right\rangle $and
electric field $\left\langle E_{x,y,z}\right\rangle $ in Ru+Ru collisions
for $b=7$ fm at $\sqrt{s_{NN}}=200$ GeV. Each figure consists of
two rows, in the first row the results from Lienard-Wiechert solution
for zero-conductivity case are presented and in the second row results
in the presence of finite conductivities are presented. In our simulation
for the case of finite conductivities, we take the values of conductivities
as $\sigma=5.8$ MeV and $\sigma_{\chi}=1.5$ MeV same as Refs. \citep{Li:2016tel,Siddique2022}.
Here we notify that $\sigma=5.8$ MeV is consistent to the lattice
QCD calculations at top temperature of QGP produced at RHIC and it
is expected that the expansion of QGP causes the decrease in $\sigma$
together with temperature of medium. Currently there is no direct
estimation towards choosing the $\sigma_{\chi}$. Since the analytic
solutions of Maxwell equations for electric and magnetic fields given
in Eqs. \ref{eq:Brphiz} and \ref{eq:Erphiz} are obtained under the
condition $\sigma\gg\sigma_{\chi}$ in Ref. \citep{Li:2016tel}, we
take $\sigma_{\chi}=1.5$ MeV as taken in the earlier work. The snapshot
for zero-conductivity case is presented at $t=0$ fm/c in the first
row while the snapshot for finite conductivities case is presented
at $t=t_{Q}$ fm/c in the second row for Ru+Ru collision at $b=7$
fm at $\sqrt{s_{NN}}=200$ GeV. We choose $t=t_{Q}$ fm/c for finite
conductivity case because we observed maximum strength of field at
this time as will be shown in upcoming Subsection \ref{subsec:Time-evolution:}.
One can notice that spatial distribution of electric and magnetic
field around $x=0$ and $y=0$ axes are symmetric when $\sigma=\sigma_{\chi}=0$.
Once we have finite conductivities then the symmetry of spatial distribution
for electric and magnetic fields is broken that is they appear symmetric
around $x=0$ axis but asymmetric around $y=0$ axis. This is due
to the presence of $\sigma_{\chi}$ in $E_{\phi}$ and $B_{r}$ in
the electric field and magnetic field respectively. More details about
broken symmetry of spatial distribution of the electric and magnetic
fields are given in Refs. \citep{Li:2016tel,Siddique2022}. We have
also checked the spatial distribution for Zr+Zr collisions at $b=7$
fm at $\sqrt{s_{NN}}=200$ GeV and found that the trends for both
cases are similar to those in Ru+Ru collisions.

\subsubsection{Centrality dependence:}

\begin{figure*}
\includegraphics[width=12cm]{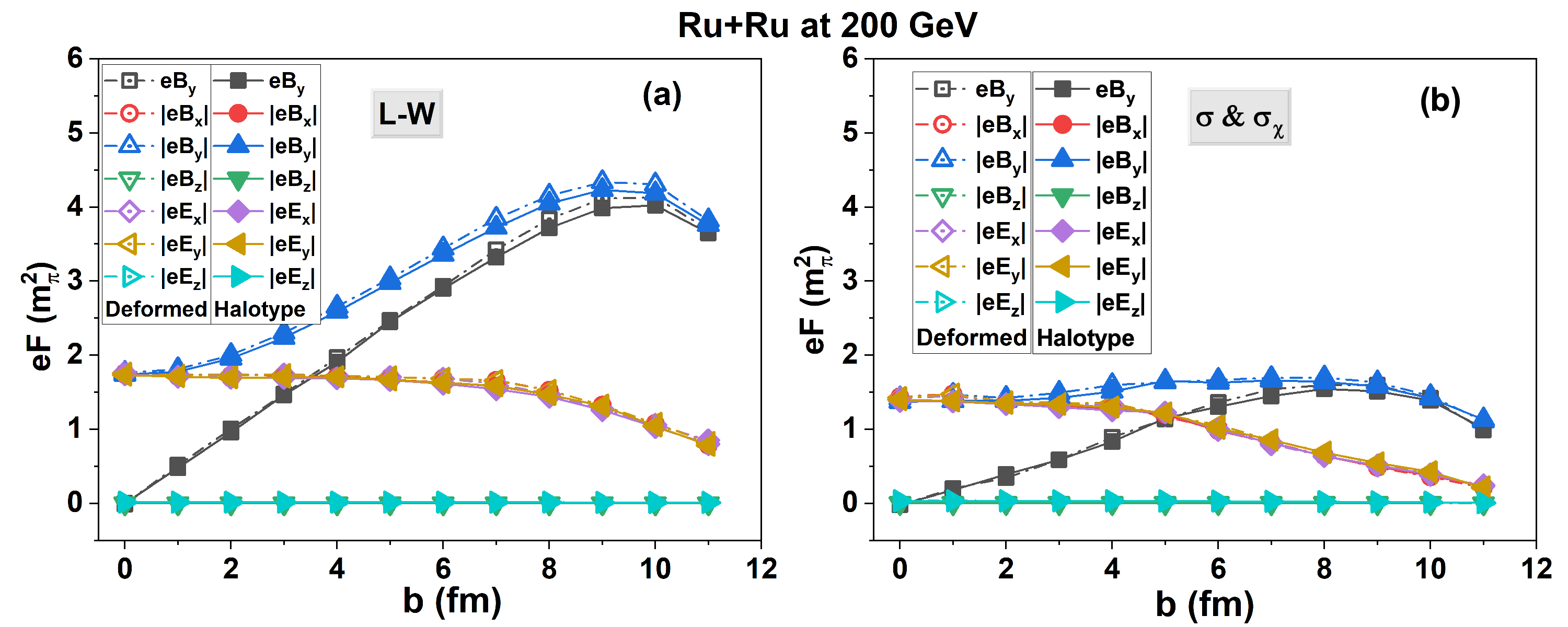}

\caption{\label{fig:Comparison-deform-halo}Comparison between $e\text{\textbf{F}}\left(m_{\pi}^{2}\right)$
for deformed (open symbols) and halo-type (closed symbols) nuclei structure
for Ru+Ru collision at $\sqrt{s}=200\text{ GeV}$.}
\end{figure*}

In this subsection we present the centrality dependence for the electric
and magnetic fields in the vacuum case and the finite conductivity
case and show comparison between them. Shown in Fig. \ref{fig:Comparison-deform-halo}
are the results for different impact parameters from central to non-central
Ru+Ru collisions with two different WS parameters for vacuum (left
panel) and finite conductivities case (right panel). Solid symbols
with solid lines represent results for halotype WS parameters while
open symbols with dashed lines are for deformed nuclei WS parameters
for Ru in Ru+R collisions. Results show very little difference when
the impact parameter is large but for small impact parameter almost
no difference is found.

\begin{figure*}
\includegraphics[width=12cm]{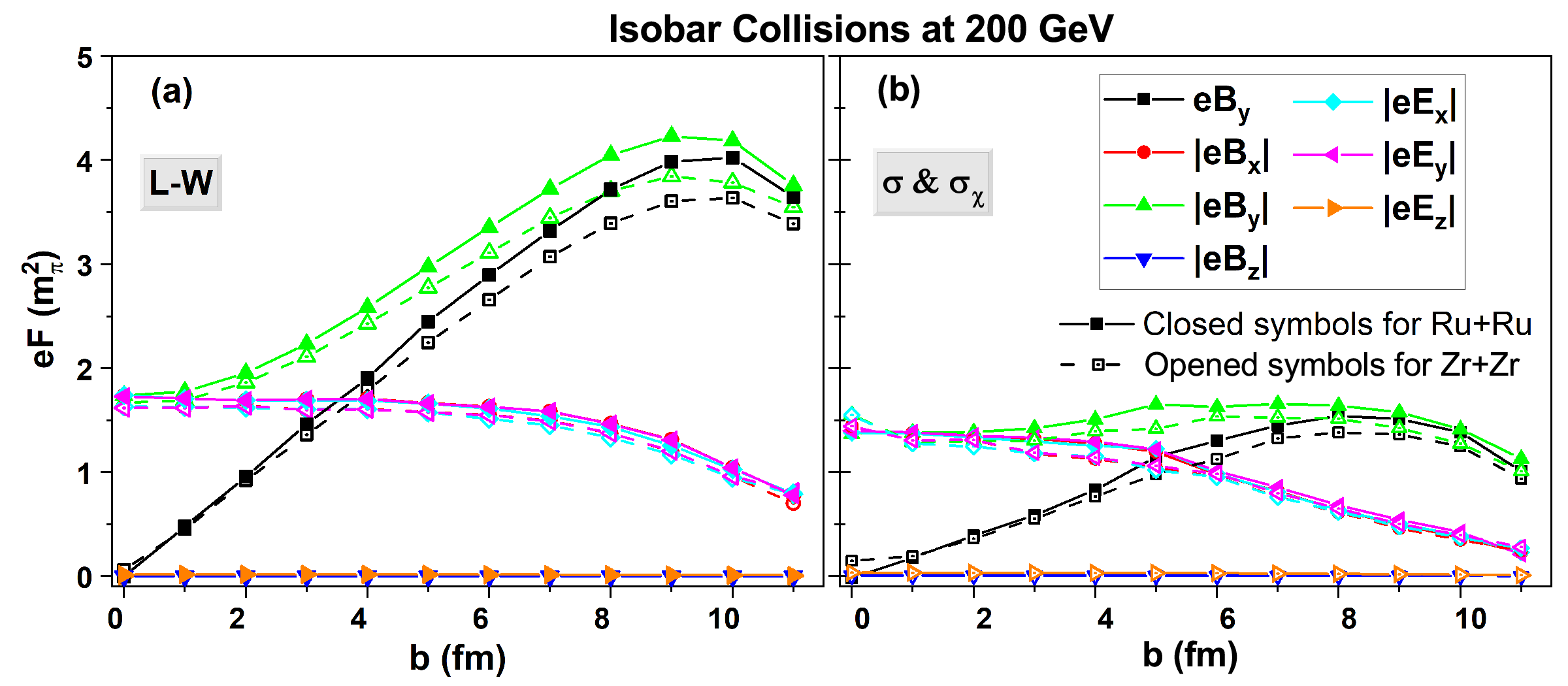}

\caption{\label{fig:impact-parameter-1} $e\text{\textbf{F}}$(in units of
$m_{\pi}^{2}$) in isobar collisions (halo-type) at $\sqrt{s_{NN}}=200$
GeV as a function of impact parameter at $t=0$ fm/c for zero conductivity
case and $t=t_{Q}$ fm/c for finite conductivities case and compared
between these two cases.}
\end{figure*}

Shown in Fig. \ref{fig:impact-parameter-1} are the results for different
impact parameters from central to non-central collisions with halotype
structure of Ru and Zr. Solid symbols with solid lines represent results
from Ru+Ru collisions while open symbols with dashed lines are results
from Zr+Zr collisions. Clearly one can see that the impact parameter
plays important role on the strength of electric and magnetic field
components. For the case of zero-conductivity as shown in Fig. (\ref{fig:impact-parameter-1}a)
the results are similar to the results of electromagnetic fields for
Au+Au and isobar collisions as in Refs. \citep{Bzdak:2011yy,Deng:2012pc}
and \citep{Zhao2019a} respectively. Moreover, the magnitudes of the
magnetic and electric fields\textbf{ $\left(e\text{\textbf{F}}\right)$}
follow $e\text{\textbf{F}}_{Au}>e\text{\textbf{F}}_{Ru}>e\text{\textbf{F}}_{Zr}$
because of the decreasing number of protons in three collision systems
respectively. Note that our simulation results show $\left|eB_{x}\right|\approx\left|eE_{x}\right|\approx\left|eE_{y}\right|$.
For the case of finite conductivities as shown in Fig. (\ref{fig:impact-parameter-1}b),
we present the impact parameter dependence of electric and magnetic
field components at time $t_{Q}$. We clearly see that the presence
of conductivities suppresses the strength significantly for $eB_{y}$
and $\left|eB_{y}\right|$, however the magnitudes for $\left|eB_{x}\right|$
and $\left|eE_{x,y}\right|$ are still comparable in the vacuum and
finite conductivity cases for small impact parameters. Longitudinal
components of electric and magnetic field $eF_{z}$ are always much
smaller and consistent to zero in comparison with transverse components
for both vacuum and finite conductivity cases.

\subsubsection{\label{subsec:Time-evolution:}Time evolution:}

\begin{figure}
\includegraphics[width=8cm]{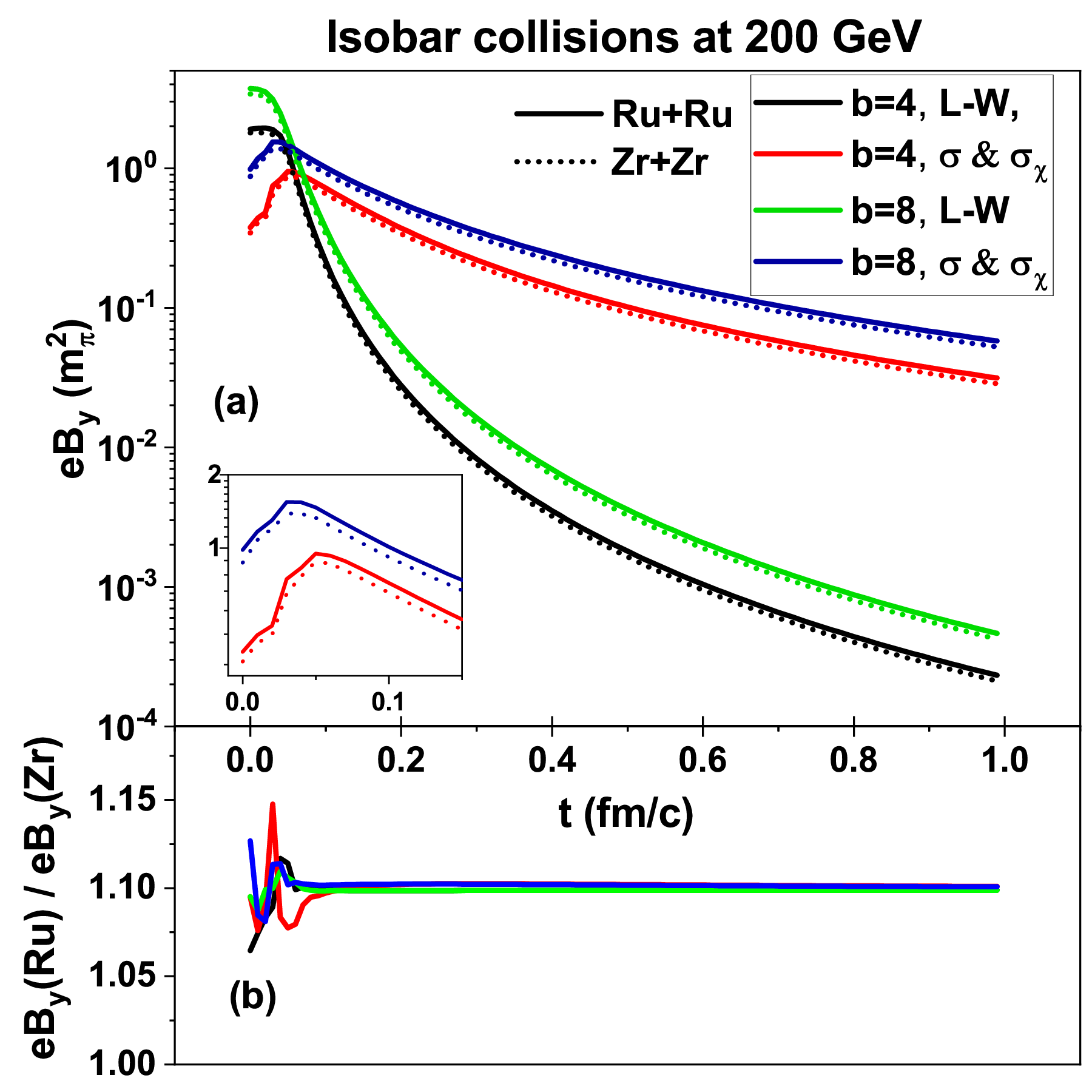}

\caption{\label{fig:Time-evolution}Time evolution of magnetic field $eB_{y}$
in isobar collisions for $b=4,8$ fm at $\sqrt{s_{NN}}=200$ GeV,
compared between zero-conductivity case and finite conductivities
case at (0,0,0).}
\end{figure}

In this subsection we show the time evolution of electric and magnetic
fields in vacuum and finite-conductivities systems and show comparison
between them. Since the dominant component is the magnetic field perpendicular
to reaction plane so in Fig. \ref{fig:Time-evolution} we show the
time evolution of $eB_{y}$ in the unit of $m_{\pi}^{2}$ for isobar
collisions in the log-scale. From the figure we can see that for both
isobar collisions although the magnitude for the vacuum case is large
as compared to the conductivity case at the beginning but the former
damps faster than the later. From the figure we also see that in case
of conductivities, time evolution has a peak and reaches its maximum
at $t_{Q}\sim0.05$ (0.03) fm/c for $b=8$ (4) fm in our calculations.
So all the results which we present in this paper for the case of
conductivities are at $t=t_{Q}$ fm/c because the maximum strength
of field is achieved at this time. We also see that simulation results
of Zr+Zr collisions are smaller as compared to Ru+Ru collisions due
to less number of protons. We also compare the ratio $eB_{y}(Ru)/eB_{y}(Zr)$
in the bottom of the figure from Ru to Zr in two cases and they are
around 1.1 consistent with the ratios of protons from Ru to Zr (Ru/Zr)
showing the difference of 10\% in magnetic field strength.

\subsection{Azimuthal angle correlation between electromagnetic field and matter
geometry}

In this work we focus on taking into account the feedback effects
from the electric and chiral magnetic conductivities and see their
effects on the azimuthally fluctuating directions of electric and
magnetic field. The effect of finite conductivities on spatial distribution
and time behavior of the electromagnetic field in isobar collisions
have been discussed in the previous subsection, we further extend
our investigation to the correlation between azimuthally fluctuating
electromagnetic fields with matter geometry characterized by participant
plane and also give their effects on EM field related observables.
So in the following subsection we give our exploration for these correlations.

\subsubsection{Magnetic field and participant plane:}

In this subsection we show the study of the correlation between azimuthal
direction of magnetic field $\left(\Psi_{B}\right)$ and participant
plane $\left(\Psi_{2}\right)$. The correlations have been studied
in the presence of finite conductivities and compared to the zero-conductivity
system. Before we give numerical results it is important to mention
that the electric and magnetic field fluctuates strongly in azimuthal
direction and magnitude on the event-by-event basis and so does the
participant plane. It is important to take into account the orientation
of $\text{\textbf{B}}$ field with the corresponding matter geometry
on event-by-event basis. Also the magnetic field induced effects which
we observe occurs along or perpendicular to the direction of magnetic
field, so it is important to determine the direction of \textbf{$\text{\textbf{B}}$}
field in accordance with experiments. By studying the correlation
between $\Psi_{B}$ and $\Psi_{2}$ we can determine the distribution
of relative angle on event-by-event basis over many events and eventually
we can give their effects on CME observable. Here we notify that usually
the participant plane $\Psi_{n}$ is rotated by $\pi/n$ as shown
in Eq. \ref{eq:psin} to serve as proxy for the event plane measured
in experiments \citep{Qin:2010pf} and also rotation is performed
for the condition of sufficiently small elliptic flow \citep{Teaney:2010vd},
however in our model we do not consider this rotation. In our study
we focus on seeing the effect of finite $\sigma$ and $\sigma_{\chi}$
on the correlation between magnetic field and initial geometry of
colliding system in Ru+Ru and Zr+Zr collisions.

\begin{figure*}
\includegraphics[width=14cm]{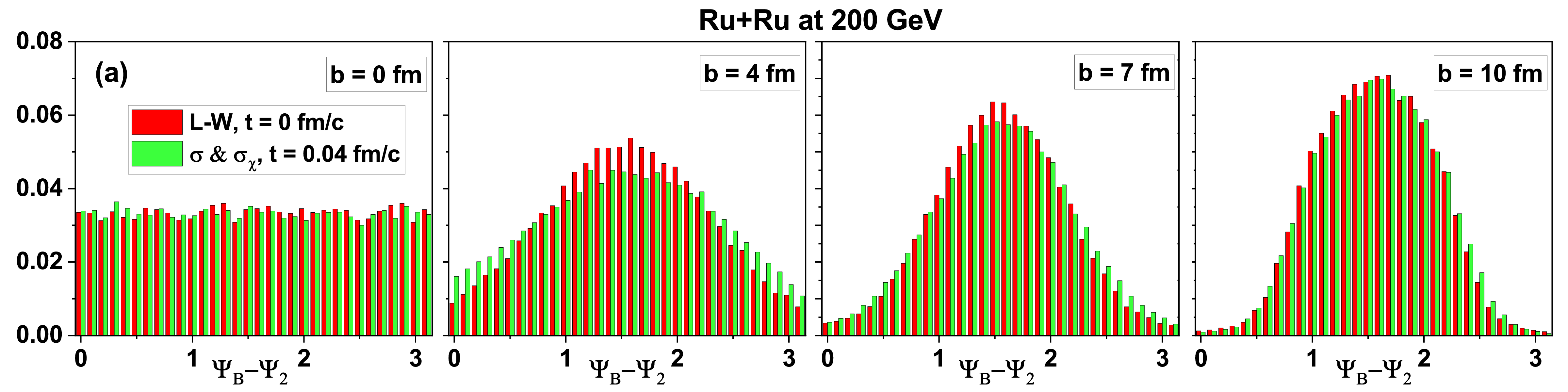}

\includegraphics[width=14cm]{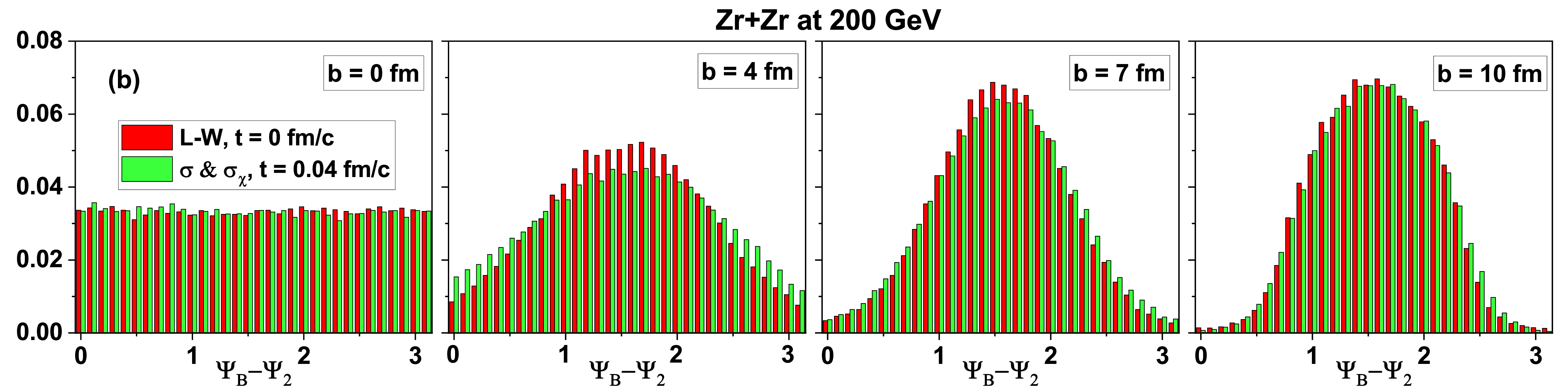}

\caption{\label{fig:Histogram-Ru}The histograms of $\Psi_{B}-\Psi_{2}$ on
event-by-event basis and compared between vacuum case (L-W) and finite
conductivity case for Impact parameter $b=0,4,7,10$ at $\boldsymbol{r}=(0,0,0)$.
Histogram results for Ru+Ru collisions are in the first row and for
Zr+Zr collisions are in the second row at $\sqrt{s_{NN}}=200$ GeV.}
\end{figure*}

In Fig. \ref{fig:Histogram-Ru} we give the histograms of $\Psi_{B}-\Psi_{2}$
on event-by-event basis for Ru+Ru collisions (first row i.e., Fig.
(\ref{fig:Histogram-Ru}a)) and Zr+Zr collisions (second row i.e.,
Fig. (\ref{fig:Histogram-Ru}b)) at $\sqrt{s_{NN}}=200$ GeV for $b=0,4,7,10$
fm. In each row of figure we compare histograms obtained from the
zero-conductivity case to the finite conductivities case. We calculate
the $\Psi_{B}$ at $\boldsymbol{r}=(0,0,0)$, and $t=0$ fm/c for
zero-conductivity case and $t=t_{Q}$ fm/c for finite conductivity
case. The histograms given in Fig. (\ref{fig:Histogram-Ru}a) for
Ru+Ru collisions at $b=0$ fm are almost uniform which represents
$\Psi_{B}$ and $\Psi_{2}$ are uncorrelated in fully overlapped collisions
in both vacuum and finite-conductivity case. However for $b>0$ fm
as shown in figure for $b=4,7,10$ fm histograms shows certain correlation
between $\Psi_{B}$ and $\Psi_{2}$ which have trends similar to Gaussian
shape with wide widths and peaking at $\pi/2$ for both zero-conductivity
and finite conductivities case. The peak of the histogram for zero-conductivity
case is larger and its width is narrower as compared to finite conductivity
case. This difference of peak and width is very minimal in peripheral
collisions at $b=10$ fm as there are very less number of participants
at this impact parameter. The similar behavior is observed in Zr+Zr
collisions as shown in the Fig (\ref{fig:Histogram-Ru}b). The Fig.
\ref{fig:Histogram-Ru} is for deformed nuclei parameters, we checked
that halotype nuclei parameters also have similar histograms.

\begin{figure*}
\includegraphics[width=14cm]{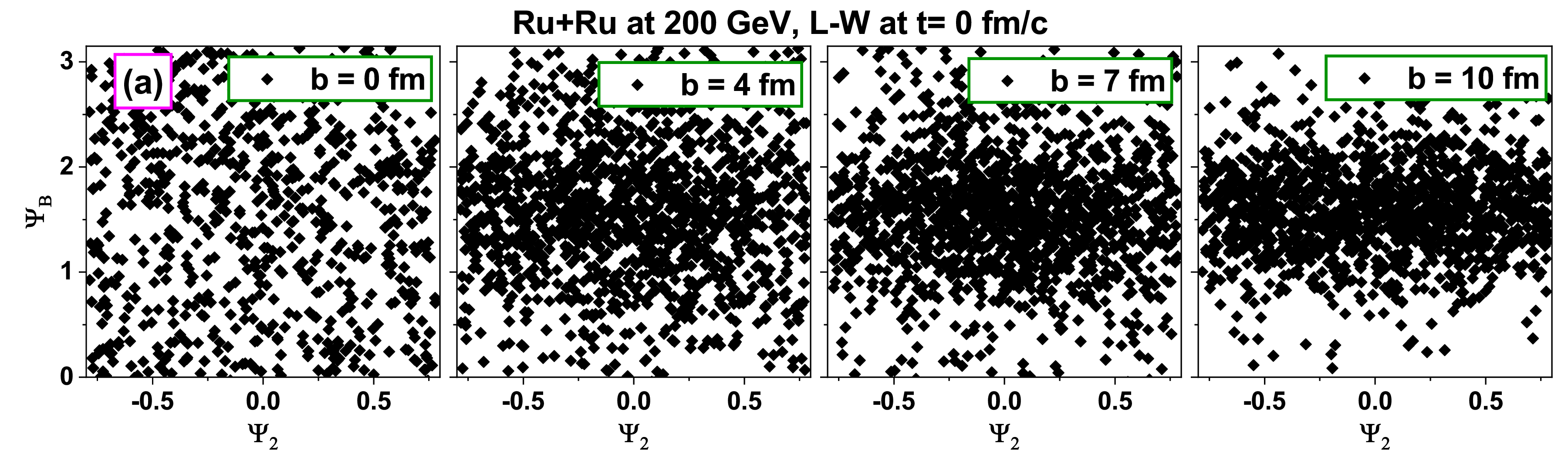}

\includegraphics[width=14cm]{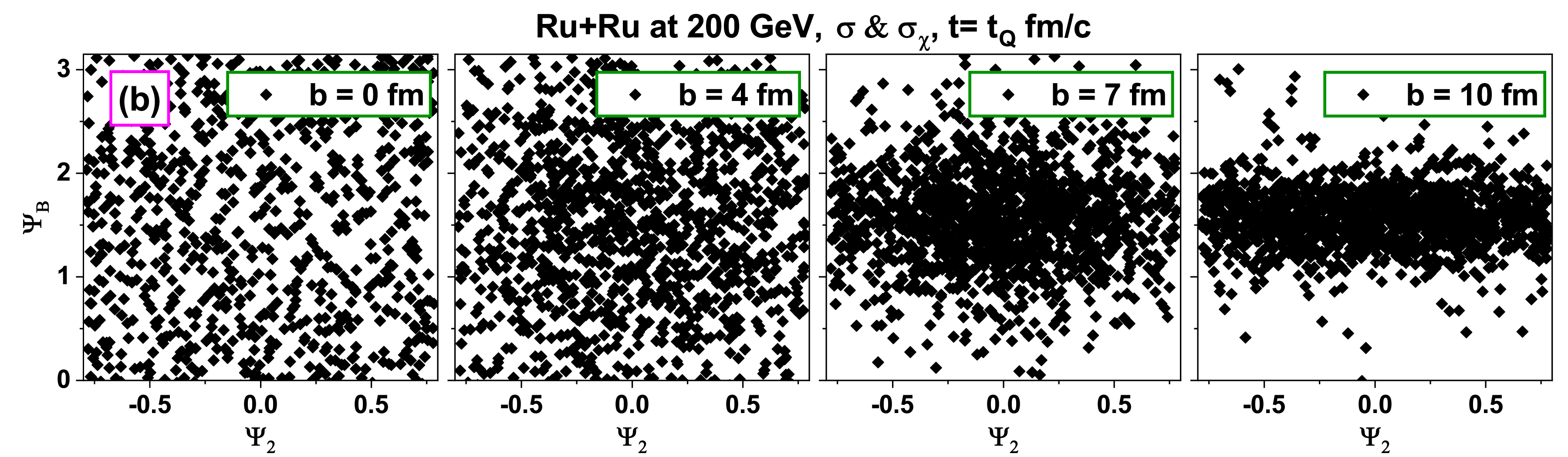}

\caption{\label{fig:psin-psi2-distribution}The scatter plots on $\Psi_{B}-\Psi_{2}$
plane in Ru+Ru collisions at $\sqrt{s_{\mathrm{NN}}}=200$~GeV for
impact parameters $b=0,4,7,10$ fm and compared between zero-conductivity
case (first row) to finite conductivities case (second row).}
\end{figure*}

Shown in Fig. \ref{fig:psin-psi2-distribution} are the results for
2D distribution plots of $\Psi_{B}$ and $\Psi_{2}$ for impact parameters
$b=0,4,7,10$ fm at $\sqrt{s_{NN}}=200$ GeV in Ru+Ru collisions.
The scatter plots obtained in zero-conducting medium are presented
in the first row (Fig. (\ref{fig:psin-psi2-distribution}a)), while
the scatter plots obtained in the conducting medium case are presented
in the second row (Fig. (\ref{fig:psin-psi2-distribution}b)). Again
for $b=0$ fm for both cases the distribution is uniform showing extremely
weak correlation in fully overlapped collisions. However for $b=4,7,10$
fm the distribution of scatter plot shows the concentration of distributions
at $\left(\Psi_{B},\Psi_{2}\right)=\left(\pi/2,0\right)$ indicates
the existence of correlation between two angles in both non-conducting
and conducting medium. We observed the spread in $\Psi_{B}$ is thinner
in finite conductivity case as compared to zero-conductivity case.
We have also checked 2D scatter plots for Zr+Zr collisions and the
behavior is similar to that found in Ru+Ru collisions. Our histogram
results and scatter plots between $\Psi_{B}$ and $\Psi_{2}$ are
similar to those reported in Refs. \citep{Bloczynski:2012en,Zhao2019a}.
The Fig. \ref{fig:psin-psi2-distribution} is for deformed nuclei
parameters, we checked that halotype nuclei parameters also have similar
scatter plots.

\begin{figure*}
\includegraphics[width=12cm]{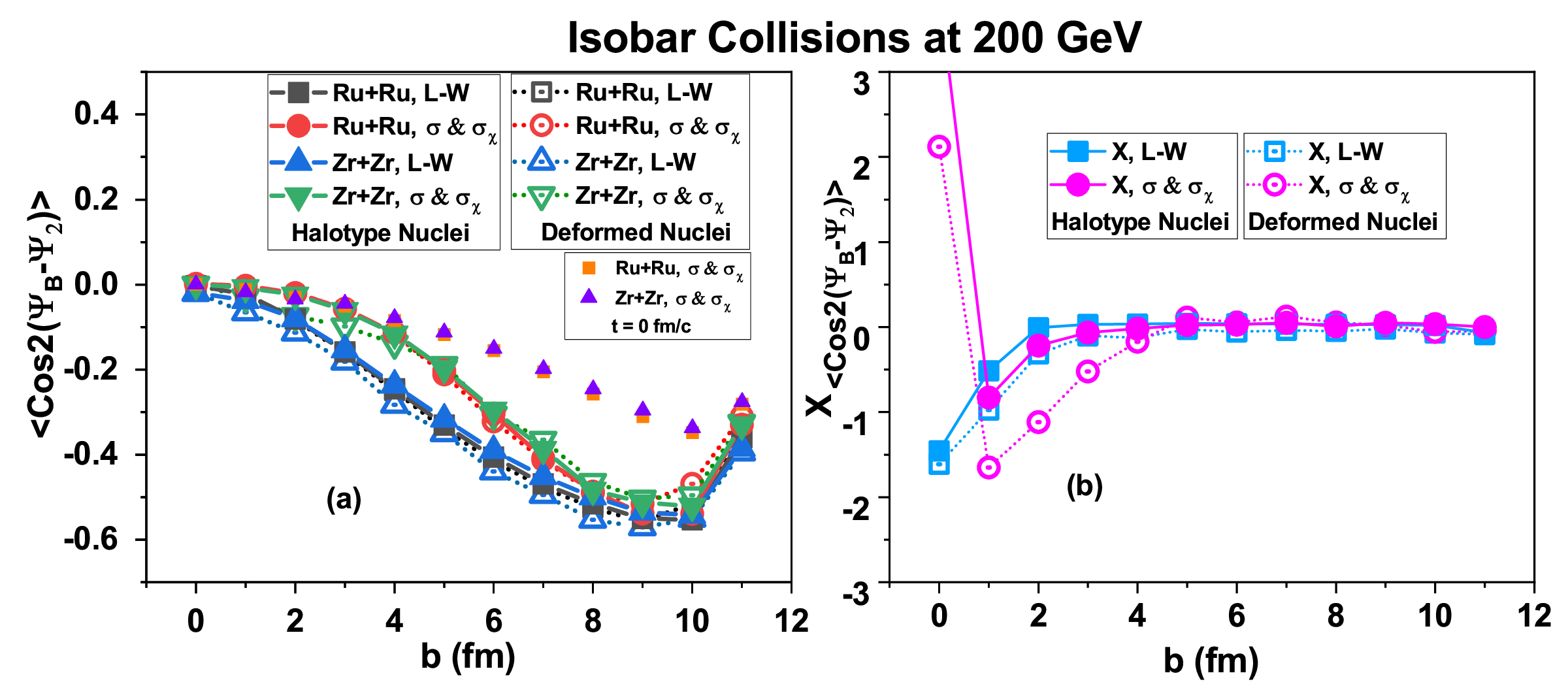}

\includegraphics[width=12cm]{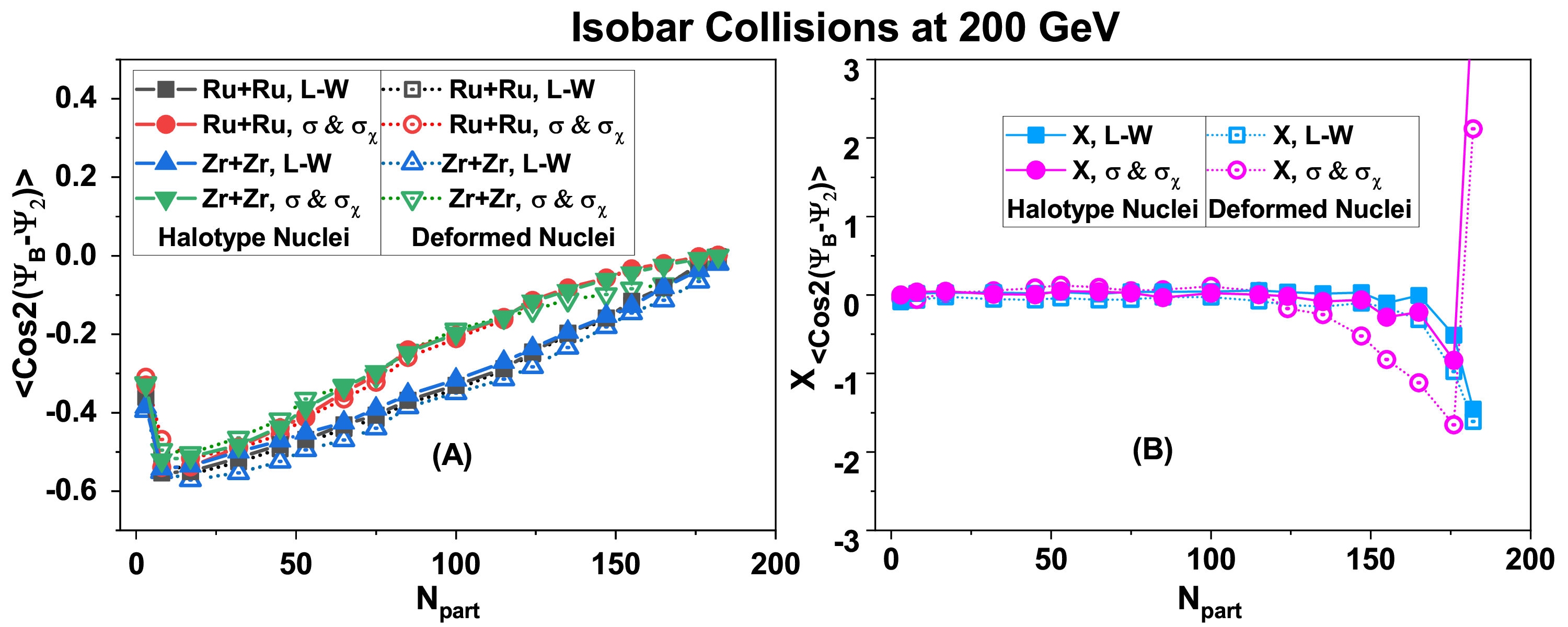}

\caption{\label{fig:cor-psib-psi2}The correlation $\left\langle \cos2\left(\Psi_{B}-\Psi_{2}\right)\right\rangle $
as a function of impact parameter $b$ in the first row ($N_{part}$
in the second row) in transverse plane respectively and their relative
ratios in isobar collisions at $\sqrt{s_{NN}}=200$ GeV, compared
between zero-conductivity case $\left(t=0\text{ fm/c}\right)$ and
finite conductivities case $\left(t=t_{Q}\right)$. Orange and purple
symbols in (a) are for finite conductivity case at $t=0$ fm/c.}
\end{figure*}

Shown in the Fig. \ref{fig:cor-psib-psi2} is the correlation between
$\Psi_{B}$ and $\Psi_{2}$ in isobar collisions as a function of
$b$ in the first row ($N_{part}$ in the second row) and compared
the correlation between zero-conductivity and finite conductivities
case. Results with solid symbol with solid lines are from halotype
nuclei while open symbol with dotted lines are from deformed nuclei
parameters. There is small difference found for both parameter settings
which can also be seen from the relative ratios in Fig. (\ref{fig:cor-psib-psi2}b).
From the figure we can also see that the correlation $\left\langle \cos2\left(\Psi_{B}-\Psi_{2}\right)\right\rangle $
depends on centrality. Since spatial distribution of magnetic fields
are symmetric around $x=0$ axis but asymmetric around $y=0$ axis
in the presence of finite conductivities, we observe their effects
on correlation as well on the transverse plane. On whole, results
are consistent with the histograms and scatter plots given in Figs.
\ref{fig:Histogram-Ru} and \ref{fig:psin-psi2-distribution} that
is for small impact parameter and very large impact parameter the
correlation is almost zero and very weak in zero-conductivity and
finite conductivities cases. However, from figure we also see that
at approximately $b=7-9$ fm, the correlation $\left\langle \cos2\left(\Psi_{B}-\Psi_{2}\right)\right\rangle $
reach their maximum value about $-0.55$ for vacuum case but the maximum
value for finite conductivities case is smaller. From figure we also
see that taking into account finite $\sigma$ and $\sigma_{\chi}$
results are quantitatively different in magnitudes of correlation
as compared to vacuum case but qualitatively consistent with the vacuum
case. The results for vacuum case are almost similar to those given
in Ref. \citep{Zhao2019a}. Quantitatively the correlation at (0,0,0)
is suppressed to 40\% in the presence of finite $\sigma$ and $\sigma_{\chi}$
in intermediate impact parameter (at $b=5$ fm) but this suppression
decreases with the increasing impact parameter (suppression of 9\%
at $b=10$ fm) for both Ru+Ru and Zr+Zr collisions. The relative ratio
at origin point have similar trends for both cases but conducting
medium case shows larger deviation from zero at smaller centralities.
We notify that we have compared the results at the time $t=t_{Q}$
fm/c because we observed maximum value for magnetic field at this
time for finite conductivities case, however we have also checked
that if we take earlier time or later time then magnitude of the correlation
is smaller. Correlation for finite conductivity case at t = 0 fm/c
is also shown by orange and magenta solid symbols in Fig. (\ref{fig:cor-psib-psi2}a)
where we can see that at t = 0 fm/c the correlation for finite conductivity
case is much smaller than the vacuum case.

\begin{figure*}
\includegraphics[width=12cm]{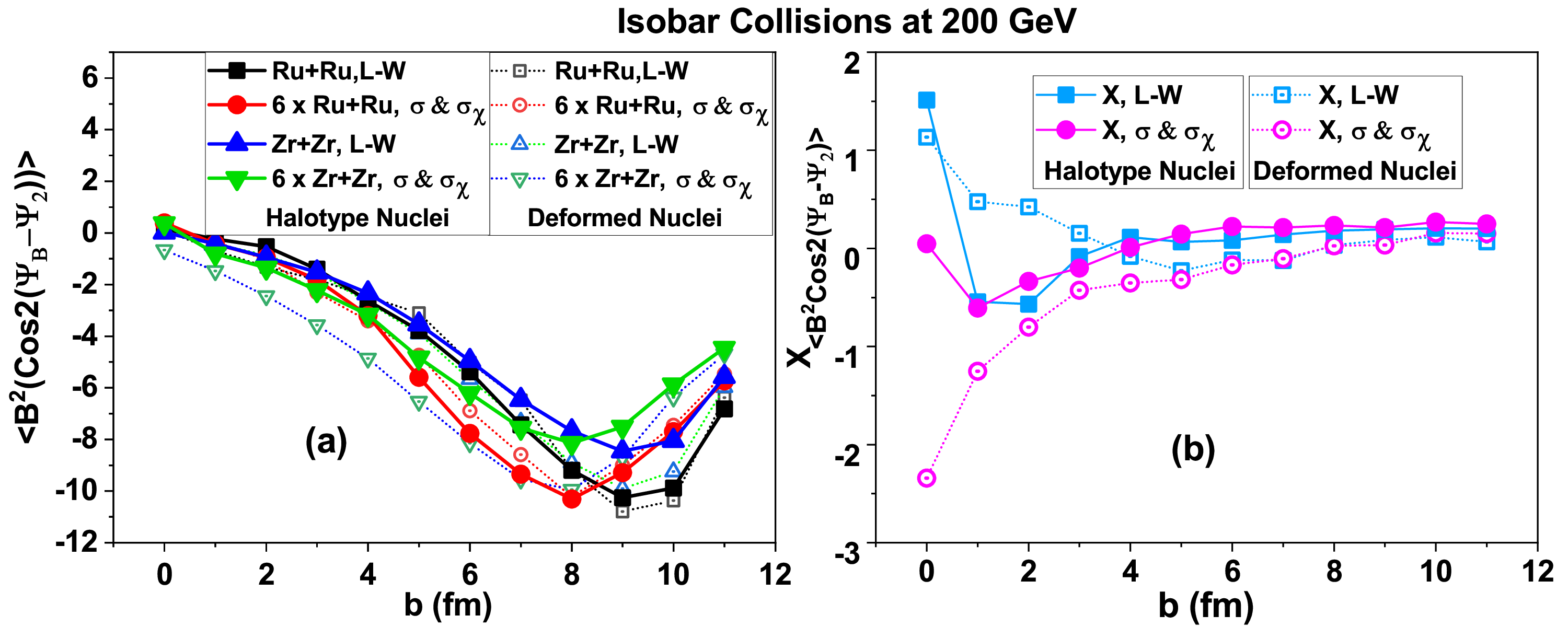}

\includegraphics[width=12cm]{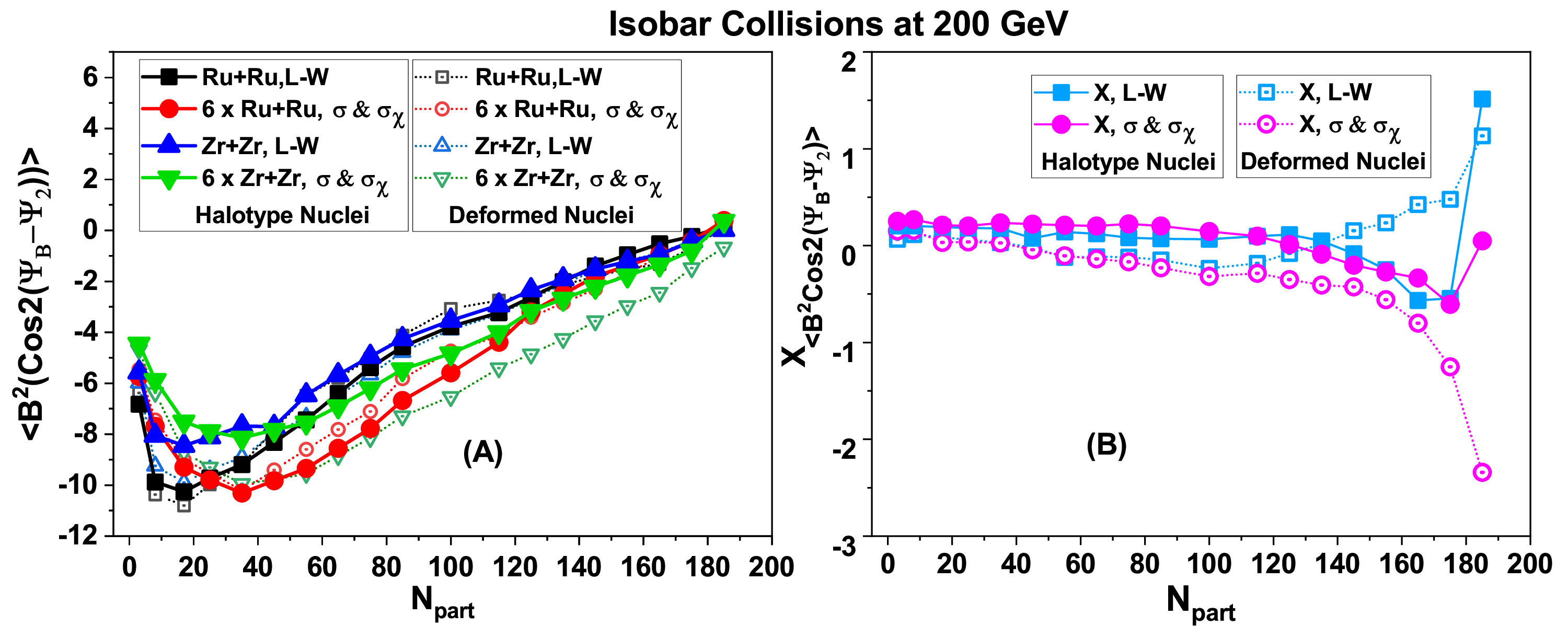}

\caption{\label{fig:-B2-psib-psi2} The correlation $\left\langle e\text{\textbf{B}}^{2}\cos2\left(\Psi_{B}-\Psi_{2}\right)\right\rangle $
as a function of impact parameter $b$ in first row ($N_{part}$ second
row) in transverse plane respectively and their relative ratios respectively
in isobar collisions at $\sqrt{s_{NN}}=200$ GeV, compared between
zero-conductivity case $\left(t=0\text{ fm/c}\right)$ and finite
conductivities case $\left(t=t_{Q}\right)$.}
\end{figure*}

According to the magnetic field induced effects, they show a directly
proportional correlation with $e\boldsymbol{\text{B}}^{2}$ along
with $\left\langle \cos2\left(\Psi_{B}-\Psi_{2}\right)\right\rangle $.
The correlator $\left\langle e\boldsymbol{\text{B}}^{2}\cos2\left(\Psi_{B}-\Psi_{2}\right)\right\rangle $
quantifies the effectiveness of magnetic field induced effect such
as generating CME signal in $\gamma$ correlator \citep{Bloczynski:2012en,Bloczynski2015}.
So in Fig. \ref{fig:-B2-psib-psi2} we show the results for $\left\langle e\boldsymbol{\text{B}}^{2}\cos2\left(\Psi_{B}-\Psi_{2}\right)\right\rangle $
in Ru+Ru and Zr+Zr collisions as a function of impact parameter $b$
in the first row ($N_{part}$ in the second row) for vacuum scenario
compared to the finite $\sigma$ and $\sigma_{\chi}$ scenario. Results
with solid symbols with solid lines represents for halotype nuclei
settings while open symbols with dashed lines represents deformed
nuclei settings. Comparing with Figs. \ref{fig:impact-parameter-1}
and \ref{fig:cor-psib-psi2} one can see that it inherits influence
from both magnetic field and correlation. We see that by considering
finite $\sigma$ and $\sigma_{\chi}$ the magnitude is decreased 6
times but qualitatively the behavior is similar to vacuum case. We
also see that maximum values of $\left\langle e\boldsymbol{\text{B}}^{2}\cos2\left(\Psi_{B}-\Psi_{2}\right)\right\rangle $
is observed at earlier centrality in case of conductivities. The sizeable
suppression can be observed for isobar collisions when compared to
$\sigma=\sigma_{\chi}=0$ case. We have also compared the relative
ratios in two different scenarios and we found that although two scenarios
differ quantitatively but qualitative trends are similar for $b>4$
fm. From the results presented in Fig. \ref{fig:-B2-psib-psi2}, we
can see that magnetic field in two different scenarios can play important
role to the charge azimuthal correlation $\Delta\gamma$ as well according
to Eq. \ref{eq:del-gama}. Our study suggest that taking into account
the feedback effects from QGP properties such finite conductivities
can also suppresses the magnetic field induced effects such as CME
signal. So when calculating the charge azimuthal correlation it is
important to incorporate QGP properties such as $\sigma$ and $\sigma_{\chi}$.

\subsubsection*{Time-Averaged Correlation}

In previous subsection we have provided comparison of the correlators
at fixed time for zero-conductivity case ($t=0$ fm/c) and finite
conductivity case ($t=t_{Q}$). Since, EM fields behavior varies with
respect to both time and space, so their impact on physical observables
should be at average level in lifespan of quark and nuclear matter.
To quantify the average effects of correlators on physical observables
we define time-averaged correlators as

\begin{equation}
\left\langle \text{\textbf{G}}\right\rangle _{t}(x)\equiv\frac{\int dt\text{\textbf{G}}(t,x)}{\int dt}\label{eq:time-averages}
\end{equation}
where \textbf{G} represents correlator $\left\langle \cos2\left(\Psi_{\text{F}}-\Psi_{2}\right)\right\rangle _{t}$
or $\left\langle e\text{\textbf{F}}^{2}\cos2\left(\Psi_{\text{F}}-\Psi_{2}\right)\right\rangle _{t}$
with F being magnetic or electric field. For numerical calculation
we discretize whole time period with discrete time points $t_{i}$
and evaluate corresponding $\text{\textbf{G}}\left(t_{i},x\right)$,
so above equation can be written into the iterative form as 
\begin{equation}
\left\langle \text{\textbf{G}}\right\rangle _{t}(x)\equiv\frac{\sum_{i}\text{\textbf{G}}\left(t_{i},x\right)\Delta t_{i}}{\sum_{i}\Delta t_{i}}\label{eq:iterative}
\end{equation}
where $\Delta t$ is time interval and we choose $\Delta t=0.02$
fm/c for our simulation. Several studies had shown that without considering
the characteristics of the medium, the maximum magnetic field occurs
in the geometric center shortly after the collision and then the magnetic
field rapidly decays with time as $t^{-3}$ in the early phase of
the evolving matter \citep{Skokov2009,Bzdak:2011yy,Voronyuk2011,Deng:2012pc,Bloczynski:2012en,Hattori2017}.
However, considering the medium feedback such as conductivities the
decay of the magnetic field is significantly slowed down, for example,
for ideal magnetohydrodynamics with infinite electric conductivity
the magnetic field decay with time as $t^{-1}$ \citep{Roy2015,Pu2016,Yan2023}.
Similar behavior is also observed and shown in Fig. \ref{fig:Time-evolution}
for decay of the magnetic field with and without taking into account
medium feedback. Also we see in Fig. \ref{fig:Time-evolution} that
for time interval $0\rightarrow1$ fm/c the magnetic field magnitude
becomes much smaller at $t=1$ fm/c than the maximum values for magnetic
fields observed soon after the collision time. However in this study
for time-averaged correlations we have considered more longer time
interval i.e, $0\rightarrow2$ fm/c. It is shown in previous subsection
that halotype nuclei and deformed nuclei structures have almost similar
results for the correlators so in this subsection we only give results
for halotype nuclei structure of Ru and Zr nuclei.

\begin{figure*}
\includegraphics[width=12cm]{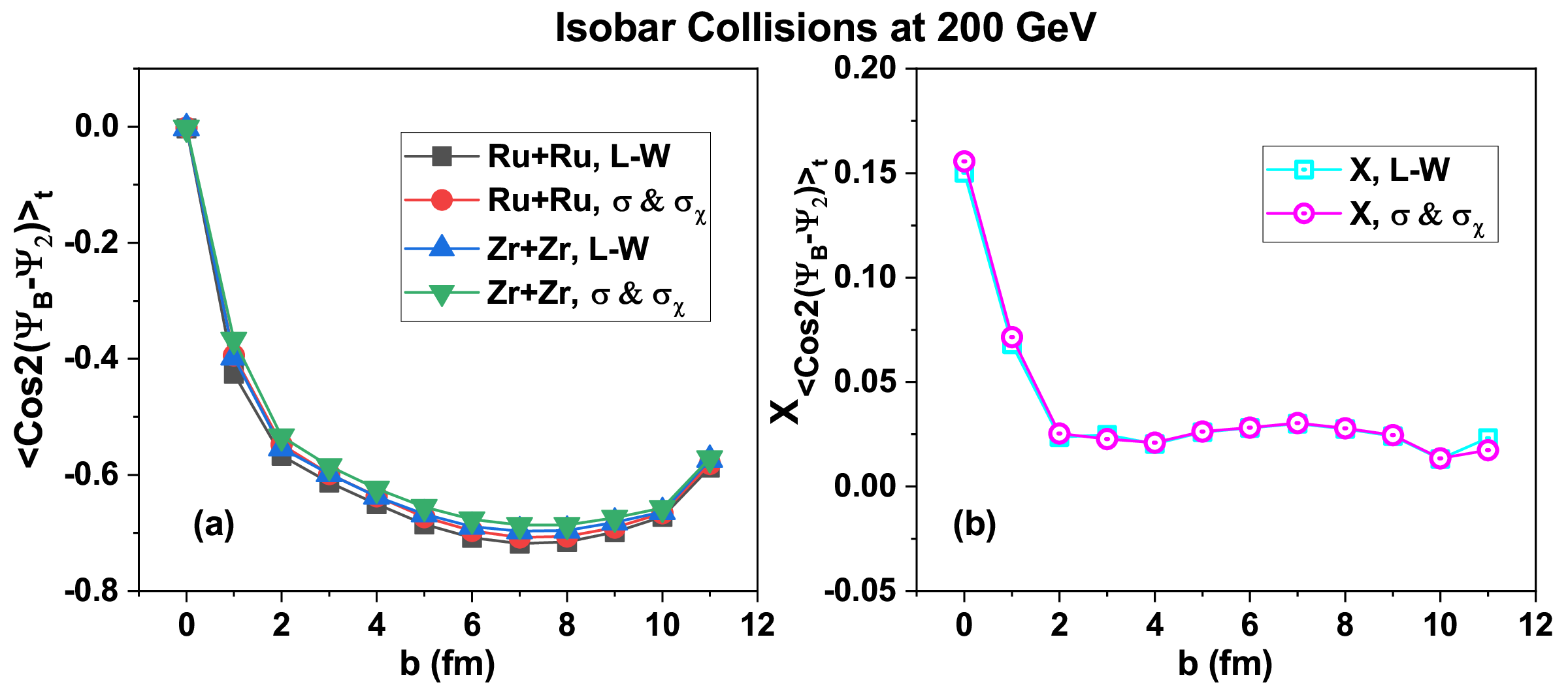}

\caption{\label{fig:-t-psib-psi2} The time-averaged correlation $\left\langle \cos2\left(\Psi_{B}-\Psi_{2}\right)\right\rangle _{t}$
as a function of impact parameter $b$ (fm) in transverse plane and
their relative ratios respectively in isobar collisions at $\sqrt{s_{NN}}=200$
GeV, comparison between zero-conductivity case and finite conductivities
case is presented.}
\end{figure*}

In Fig. \ref{fig:-t-psib-psi2}, we give the time-averaged correlation
$\left\langle \cos2\left(\Psi_{B}-\Psi_{2}\right)\right\rangle _{t}$
as a function of impact parameter $b$ fm. From the figure we can
also see that time-averaged correlation depends on different centralities.
Time-averaged correlation for zero conductivity case and finite conductivity
case shows similar behaviors, however for finite conductivities case
the magnitudes are little smaller than the zero conductivity case.
From figure we see that for $b=0$ fm time averaged correlation is
almost zero which is consistent to histogram for $b=0$ fm. The time-averaged
correlation increases with increase in impact parameter until it reaches
maximum value for $b=7-9$ fm, and then it decrease again for very
large impact parameter. Relative ratios shown in the of Fig. (\ref{fig:-t-psib-psi2}b)
have similar trends for zero conductivity case and finite conductivities
case. From figure, we see a small difference in magnitude of time-averaged
correlation when taking into account finite $\sigma$ and $\sigma_{\chi}$,f
and also qualitative behavior is consistent with the vacuum case.

\begin{figure*}
\includegraphics[width=12cm]{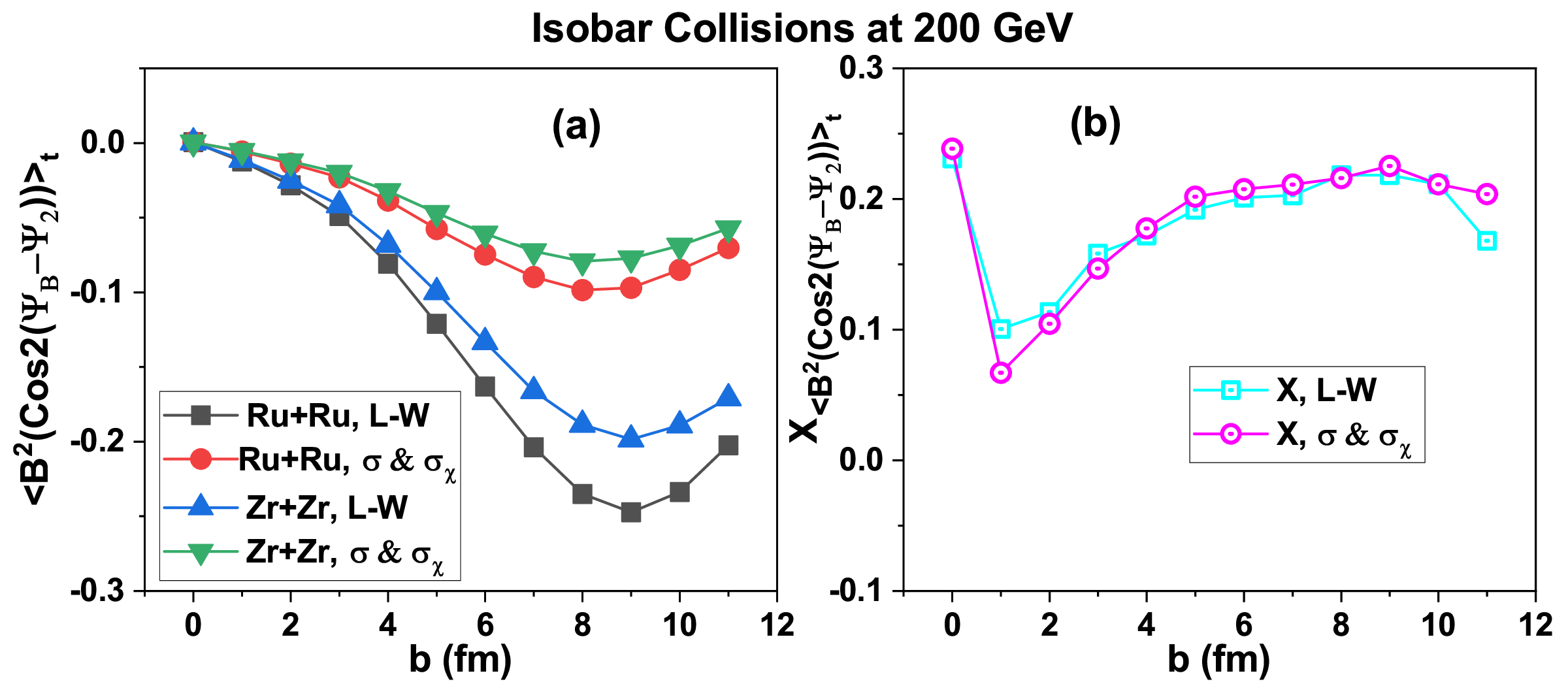}

\caption{\label{fig:-t-B2-psib-psi2} The time-averaged correlation $\left\langle e\text{\textbf{B}}^{2}\cos2\left(\Psi_{B}-\Psi_{2}\right)\right\rangle _{t}$
as a function of impact parameter $b$ (fm) in transverse plane and
their relative ratios in isobar collisions at $\sqrt{s_{NN}}=200$
GeV, comparison between zero-conductivity case and finite conductivities
case is presented.}
\end{figure*}

In Fig. \ref{fig:-t-B2-psib-psi2}, we show time-averaged correlation
$\left\langle e\text{\textbf{B}}^{2}\cos2\left(\Psi_{B}-\Psi_{2}\right)\right\rangle _{t}$
as a function of impact parameter $b$ fm. We show the results for
Ru+Ru and Zr+Zr collisions for vacuum scenario compared to finite
$\sigma$ and $\sigma_{\chi}$ scenario. The results show that they
inherits influence from both squared magnetic field as well as from
correlation $\cos2\left(\Psi_{B}-\Psi_{2}\right)$ on time averaged
level. We also see that for time interval $0-2$ fm/c time-averaged
correlation in the presence of finite $\sigma$ and $\sigma_{\chi}$
is roughly 2.5 times smaller than time-averaged correlation for zero-conductivity
case. Although the magnetic field in the presence of conductivities
have larger magnitude at later time of evolution when compared with
zero conductivity case but the interplay between squared magnetic
field $e\text{\textbf{B}}^{2}$ and correlation $\cos2\left(\Psi_{B}-\Psi_{2}\right)$
becomes smaller when taking time-averaged in interval. Also at initial
time $\left\langle e\text{\textbf{B}}^{2}\cos2\left(\Psi_{B}-\Psi_{2}\right)\right\rangle $
is much larger then the finite conductivity case so even after calculating
time-averaged correlation the magnitude stays larger. On whole qualitative
behavior for time-averaged correlation is similar to that observed
in Fig. \ref{fig:-B2-psib-psi2}. The time-averaged correlation $\left\langle e\text{\textbf{B}}^{2}\cos2\left(\Psi_{B}-\Psi_{2}\right)\right\rangle _{t}$
also suggests that by taking into account the medium feedback effects
from QGP properties such finite conductivities can also suppresses
the magnetic field induced effects such as CME effects.

\subsubsection{Electric field and participant plane}

\begin{figure*}
\includegraphics[width=9cm]{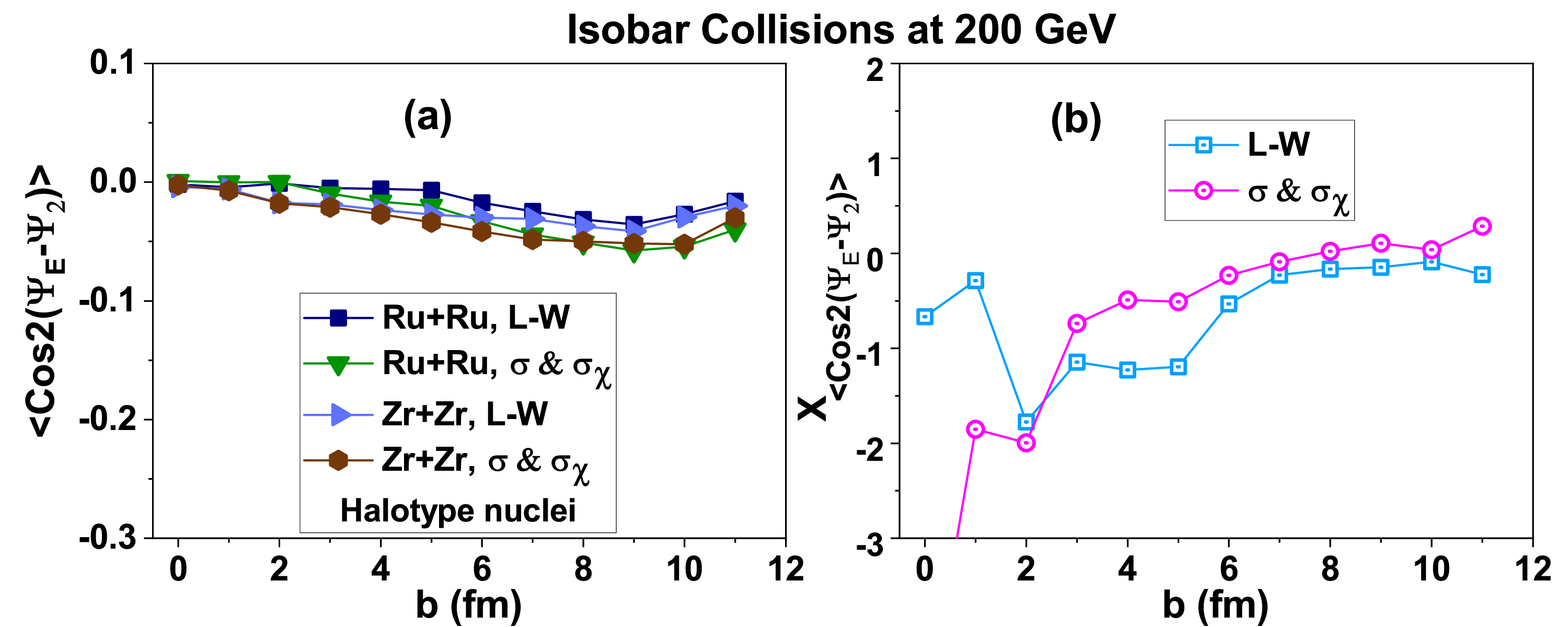}\includegraphics[width=9cm]{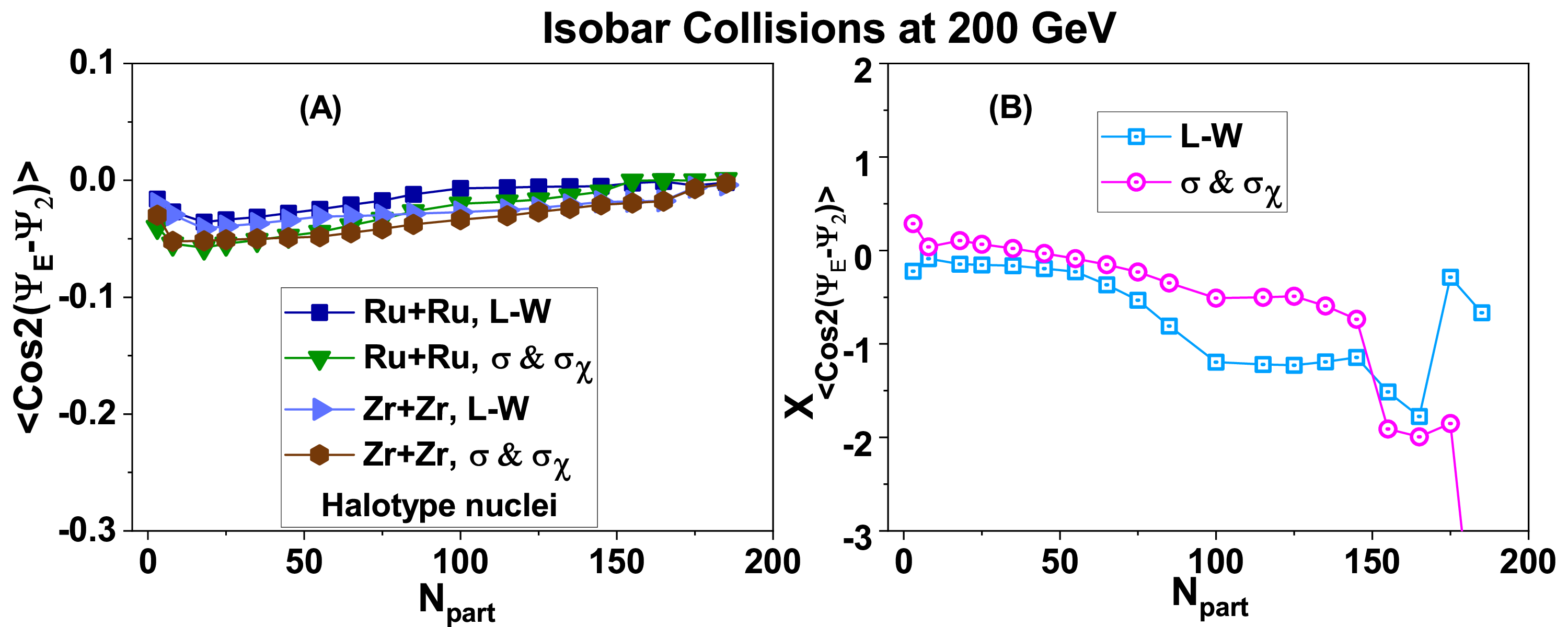}

\includegraphics[width=9cm]{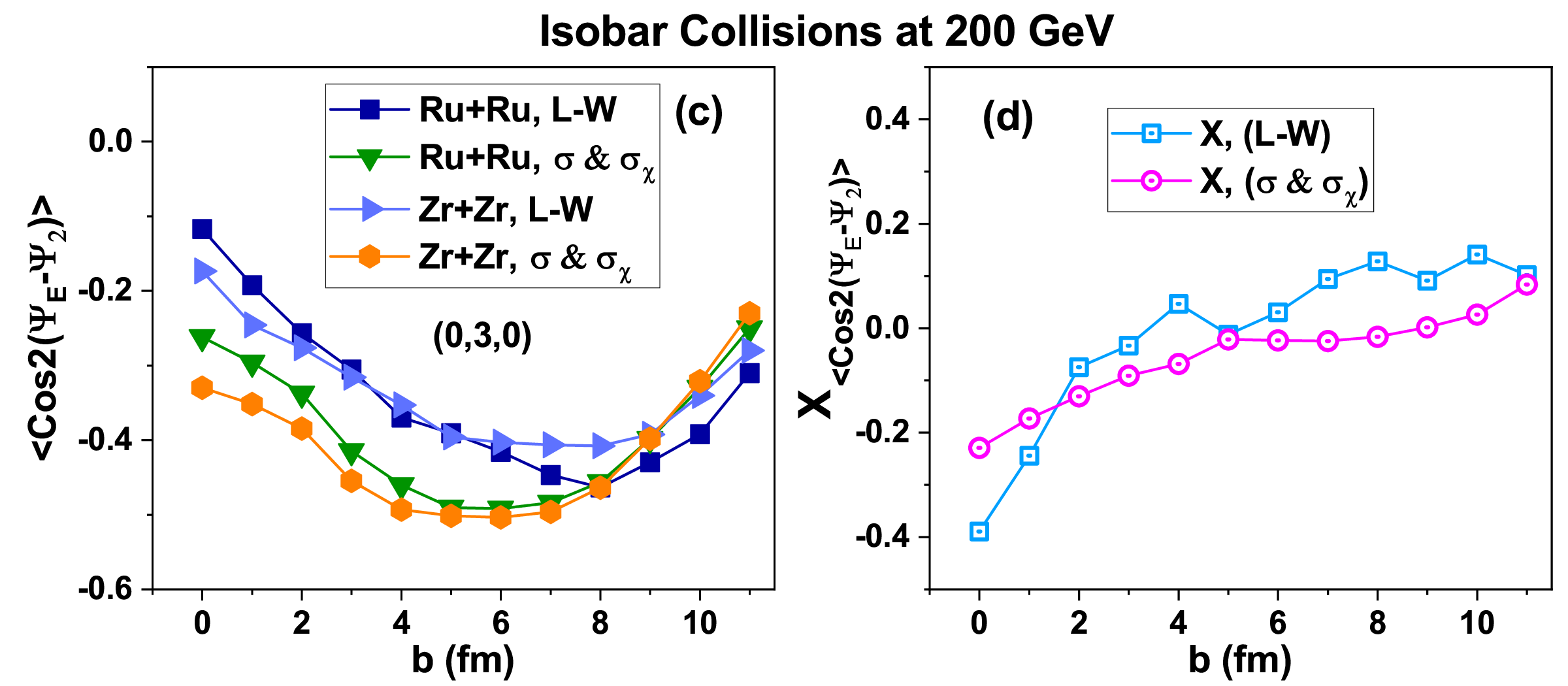}\includegraphics[width=9cm]{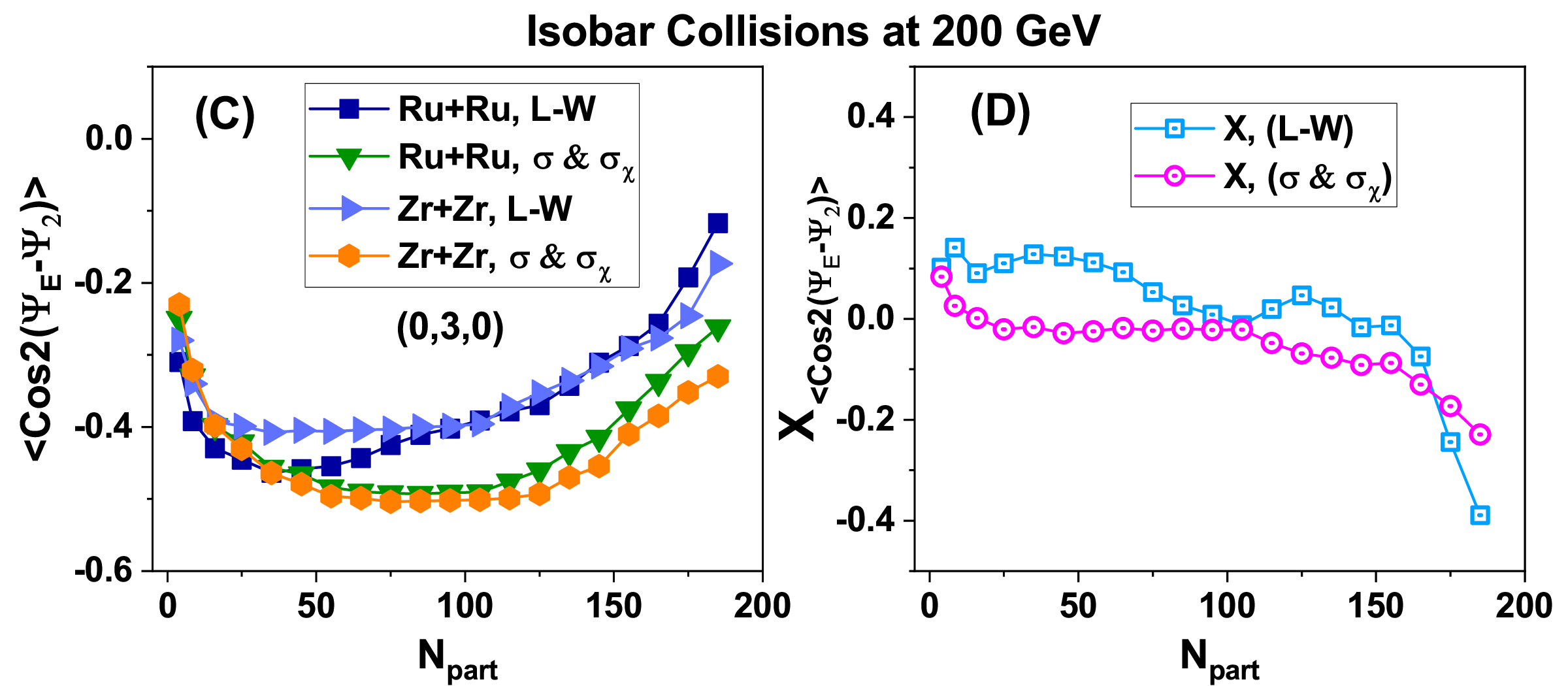}

\caption{\label{fig:-psie-psi2} The correlation $\left\langle \cos2\left(\Psi_{E}-\Psi_{2}\right)\right\rangle $
as a function of impact parameter $b$ at two different positions
in transverse plane that is $\boldsymbol{r}=(0,0,0)$ and $\boldsymbol{r}=(0,3,0)$
in first and second row respectively, and their relative ratios 
in isobar collisions at $\sqrt{s_{NN}}=200$ GeV, comparison between
zero-conductivity case $\left(t=0\text{ fm/c}\right)$ and finite
conductivities case $\left(t=t_{Q}\right)$ is presented.}
\end{figure*}

In this subsection we give the brief study of the correlation between
azimuthal direction of electric field $\left(\Psi_{E}\right)$ and
participant plane $\left(\Psi_{2}\right)$ for completeness. As we
have noticed in Fig. \ref{fig:impact-parameter-1} that the electric
field can also be comparably strong with the magnetic field. Possible
charge distribution induced by strong electric field is an example
of this. Similar to magnetic field we study the correlation $\left\langle \cos2\left(\Psi_{E}-\Psi_{2}\right)\right\rangle $.
In Fig. \ref{fig:-psie-psi2}, we show the results as a function of
$b$ ($N_{part}$) in Ru+Ru and Zr+Zr collisions at $\sqrt{s_{NN}}=200$
GeV for vacuum and conducting medium case at points (0,0,0) in the
first row and (0,3,0) in the second row. As we noticed in previous
section that halotype and deformed nuclei parameters have little difference
on correlations and their relative ratios in our setup so in first
row of Fig. \ref{fig:-psie-psi2} results are obtained by using halotype
nuclei parameters and results in second row are obtained by using
deformed nuclei parameters. In Fig. (\ref{fig:-psie-psi2}a), we calculate
$\left\langle \cos2\left(\Psi_{E}-\Psi_{2}\right)\right\rangle $
as function of $b$ and we can see that the correlation is very weak
for Ru+Ru and Zr+Zr collisions for all impact parameters. The relative
ratios shown in Fig. (\ref{fig:-psie-psi2}b) for both vacuum and
medium case have similar trends for $b>3$ fm $\left(N_{part}<160\right)$.
However, this weak correlation can be understood together with the
spatial distributions of electric field components given in Fig. \ref{fig:spatial-E}
where we can see that at (0,0,0) electric field is very weak. The
results shown in the second row of the Fig. \ref{fig:-psie-psi2}
which corresponds to point (0,3,0), we see the sizeable correlation
for zero-conductivity limit and finite conductivities case. We also
observe enhancement of correlation in the presence of finite $\sigma$
and $\sigma_{\chi}$ for small centralities. The relative ratios measured
at this point for the two cases have shown similar trends for centrality
dependence. So, we see that two scenarios behave similar qualitatively
but differ in magnitude quantitatively.

\begin{figure*}
\includegraphics[width=9cm]{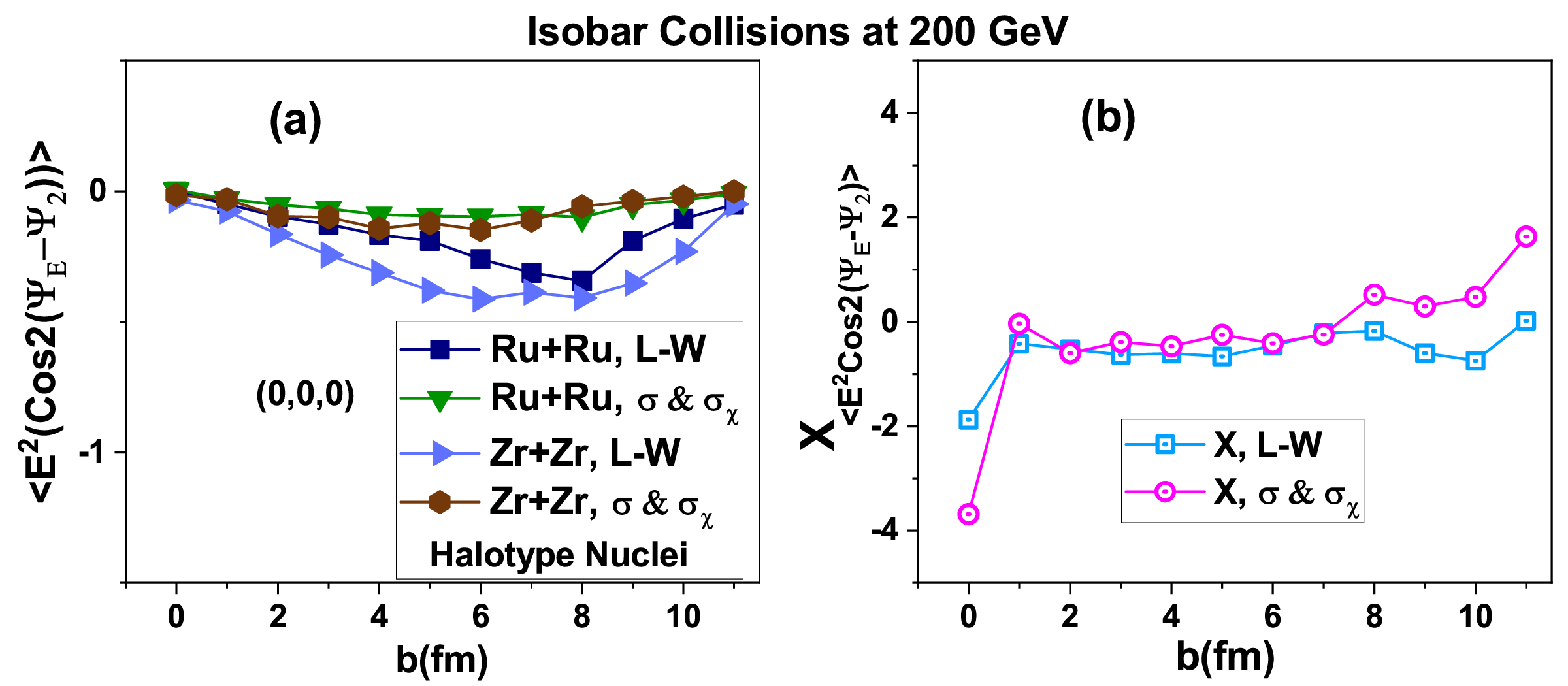}\includegraphics[width=9cm]{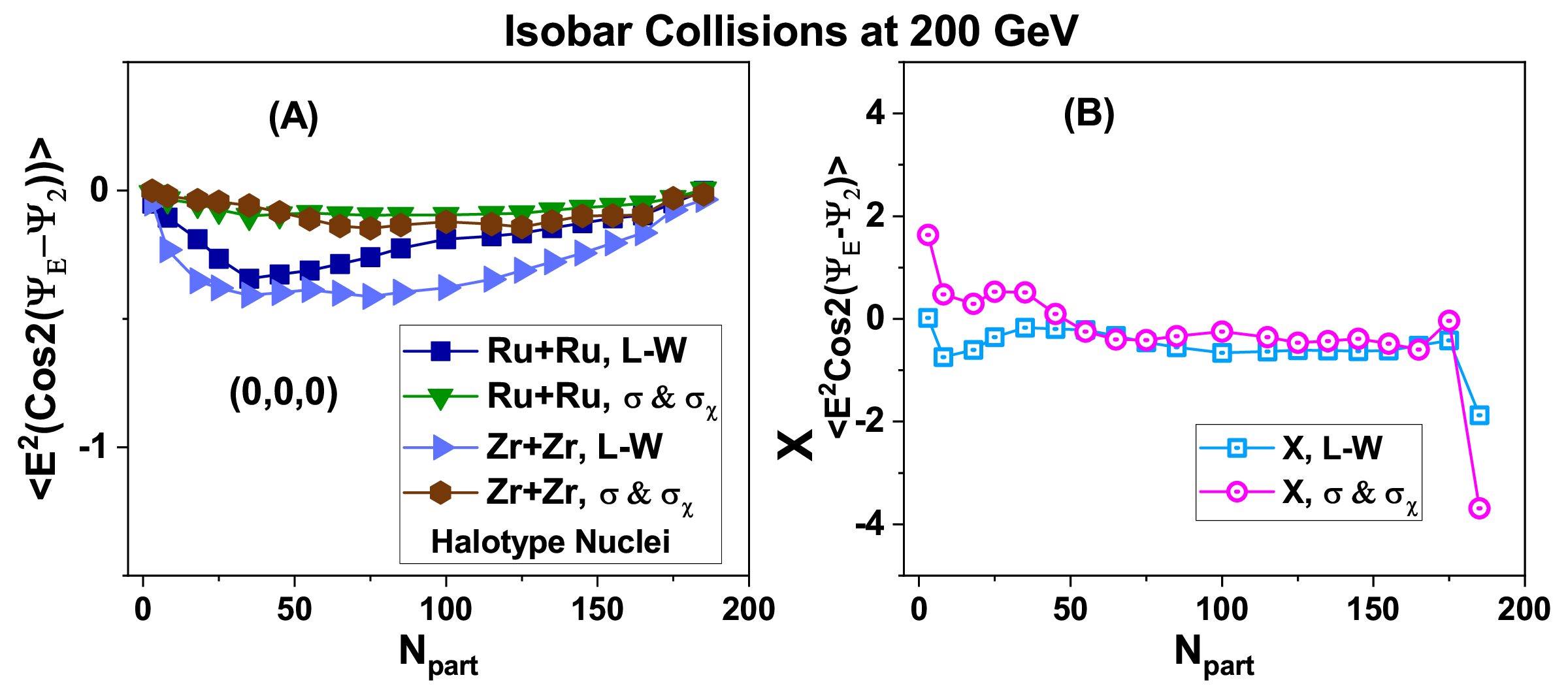}

\includegraphics[width=9cm]{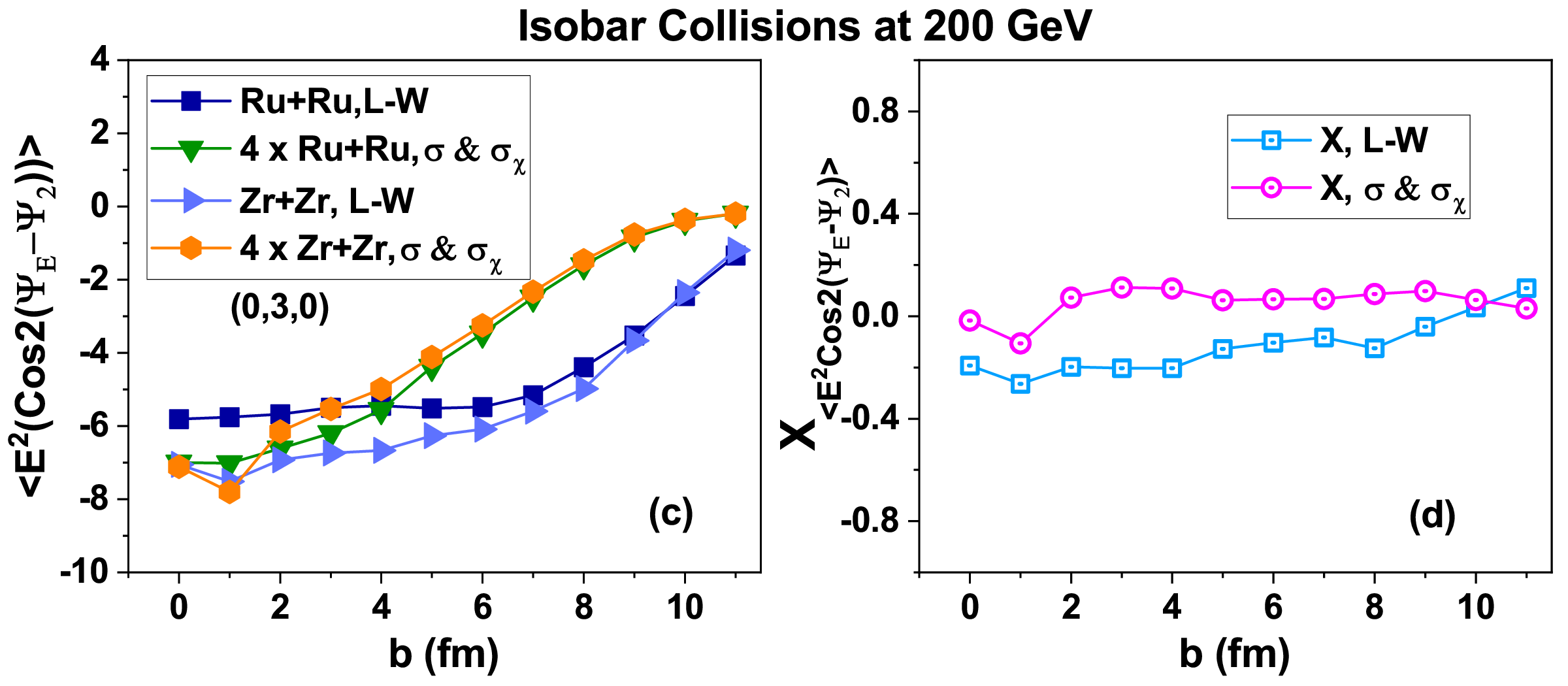}\includegraphics[width=9cm]{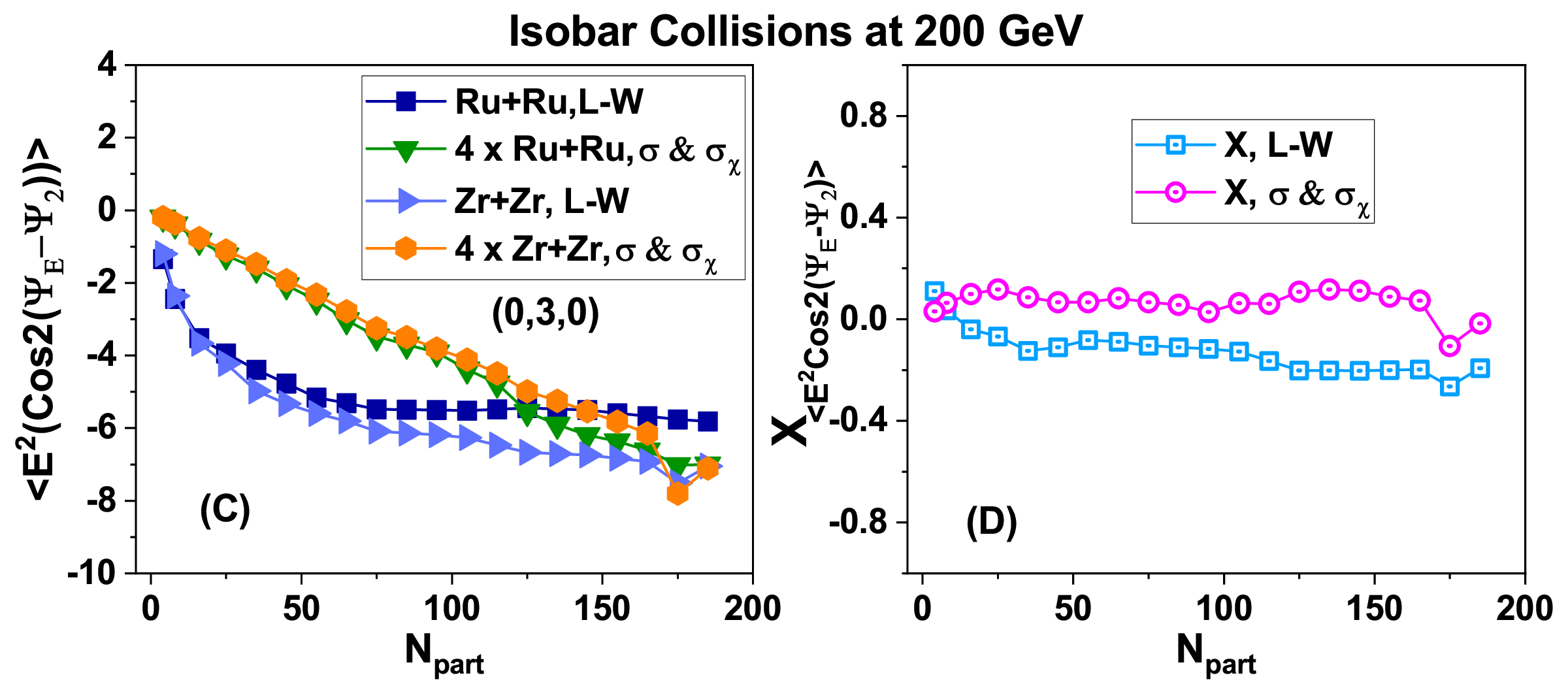}

\caption{\label{fig:-E2pise-psi2} The correlation $\left\langle e\text{\textbf{E}}^{2}\cos2\left(\Psi_{E}-\Psi_{2}\right)\right\rangle $
as a function of impact parameter $b$ at two different positions
in transverse plane that is $\boldsymbol{r}=(0,0,0)$ and $\boldsymbol{r}=(0,3,0)$
in first and second row respectively, and their relative ratios 
in isobar collisions at $\sqrt{s_{NN}}=200$ GeV, comparison between
zero-conductivity case $\left(t=0\text{ fm/c}\right)$ and finite
conductivities case $\left(t=t_{Q}\right)$ is presented.}
\end{figure*}

Similar to the magnetic field, the observable quantity related to
electric field can also be proportional to $e\text{\textbf{E}}^{2}$
and $\left\langle \cos2\left(\Psi_{E}-\Psi_{2}\right)\right\rangle $
so, in Fig. \ref{fig:-E2pise-psi2}, we calculate the correlation
$\left\langle e\text{\textbf{E}}^{2}\cos2\left(\Psi_{E}-\Psi_{2}\right)\right\rangle $
as function of $b$ ($N_{part}$) for Ru+Ru and Zr+Zr collisions at
$\sqrt{s_{NN}}=200$ GeV. We compared the results for vacuum case
with the finite conductivity case at point (0,0,0) in the first row
and at point (0,3,0) in the second row of figure. We see that it is
very weak at (0,0,0) because the strength of electric field is also
very weak at the origin point. The relative ratio at this point shows
qualitatively almost similar trend for $b$ fm $\left(N_{part}\right)$
dependence. However, we notice observable effect at point (0,3,0)
which can be understood together with the spatial distribution the
electric field at this point. From the figure we also notice that
the introduction of finite $\sigma$ and $\sigma_{\chi}$ in system
do affect the strength (4 times smaller) of correlation quantitatively
however the qualitative picture is somewhat similar to the vacuum
case. Relative ratios show different behavior at (0,3,0), for the
case of finite conductivities relative ratios are near zero however
for case of zero conductivity the relative ratio differs from zero.
The results shown in figure suggest that while calculating observable
quantity related to electric field, it is important to take into account
feedback effects from the medium properties.

\section{Summary and outlook}

\label{sec:summary}

In this study, we have conducted a systematic investigation on the
effects of electric $\left(\sigma\right)$ and chiral magnetic $\left(\sigma_{\chi}\right)$
conductivities on the impact parameter and space-time behaviors of
electromagnetic fields generated in high-energy isobar collisions.
Our results show that in transverse plane partially broken symmetry
is observed for electric and magnetic fields in isobar collisions
in the presence of finite $\sigma$ and $\sigma_{\chi}$, consistent
with those observed in Au+Au collisions. Although the magnitude is
smaller for the case of a conducting medium at earlier time but the
lifetime of fields is much longer when compared to zero-conductivity
limit (Lienard-Wiechert). We also confirm that magnetic fields differ
by 10\% even in the case of finite conductivities.

We also performed a detailed study on the azimuthal fluctuation of
electromagnetic fields $(e\text{\textbf{F}})$ in the presence of
conductivities and studied their correlations with initial matter
geometry i.e., $\left\langle \cos2\left(\Psi_{F}-\Psi_{2}\right)\right\rangle $
and $\left\langle e\boldsymbol{\text{F}}^{2}\cos2\left(\Psi_{F}-\Psi_{2}\right)\right\rangle $
on event-by-event basis. Comparison of the correlations has been presented
between the vacuum (L-W) and finite conductivities cases in this study.
We see a sizeable suppression of the correlations for fixed time and
time-averaged correlations in the presence of conducting medium, which
reflects the importance of taking into account the medium properties
such as conductivities while calculating the experimentally measurable
observable such as $\Delta\gamma$ because quantities related to EM
fields inherit influence from both the strength and the correlation
between EM field direction and initial geometry. While the relative
ratios measured in this paper show similar trends for centrality dependence
but large deviations are observed from central to mid-central collisions.
We also take into account the two different WS nuclei parameters namely
deformed nuclei and halotype nuclei and although they show difference
in results but our calculation shows that the difference is not large
and also qualitative behavior studied by relative ratios is similar
for large centralities. 

While in this work we have provided a quantitative understanding of
the influence of finite $\sigma$ and $\sigma_{\chi}$ on the electromagnetic
fields as well as their azimuthal direction correlation with the initial
geometry characterized by the participant plane in isobar collisions,
it can be improved in several directions. It is needful to integrate
these electromagnetic fields into a transport model or hydrodynamic
model, from which one may draw a more solid conclusion about the consequent
effects. In this study, we have only considered finite values for
$\sigma$ and $\sigma_{\chi}$, but the values should change with
the dynamical expansion in QGP. While in this work we only considered
the participant plane one may also consider the spectator plane to
see their correlation in the presence of conductivities. In future,
it will be interesting to study the correlation between the fluctuations
of electromagnetic anomaly $(\text{\textbf{E}}\cdot\text{\textbf{B}})$
and $n$th harmonic participant(or spectator) plane in the presence
of conductivities. In an upcoming effort, we will extend our studies
to these directions.
\begin{acknowledgments}
We are grateful to University of Chinese Academy of sciences for providing
platform to carry out simulations. Thanks to X.G. Huang and G.L Ma
for discussions during the conference Chirality, Vorticity and Magnetic
Field 2023. Thanks to Q.Wang, H. Mei, S.S. Cao, X.L. Sheng, D. Jian,
H. Xu and A. Huang for helpful discussions. This work is supported by 
the RFIS-NSFC under Grant number 12350410364 and Ministry of Science and Technology (MOST) of China under Grant 
number QN2023205001L. 
\end{acknowledgments}

\bibliographystyle{h-physrev5}
\bibliography{SCrefs}

\begin{thebibliography}{10}

\bibitem{Bzdak:2011yy}
A.~Bzdak and V.~Skokov,
\newblock Phys. Lett. B {\bf 710}, 171 (2012), arXiv:1111.1949.

\bibitem{Deng:2012pc}
W.-T. Deng and X.-G. Huang,
\newblock Phys. Rev. C {\bf 85}, 044907 (2012), arXiv:1201.5108.

\bibitem{Voronyuk2011}
V.~Voronyuk {\em et~al.},
\newblock Phys. Rev. C {\bf 83}, 054911 (2011), arXiv:1103.4239.

\bibitem{Skokov2009}
V.~Skokov, A.~Y. Illarionov, and V.~Toneev,
\newblock Int. J. Mod. Phys. A {\bf 24}, 5925 (2009), arXiv:0907.1396.

\bibitem{Zhong:2014cda}
Y.~Zhong, C.-B. Yang, X.~Cai, and S.-Q. Feng,
\newblock Adv. High Energy Phys. {\bf 2014}, 193039 (2014), arXiv:1408.5694.

\bibitem{Kharzeev:2007jp}
D.~E. Kharzeev, L.~D. McLerran, and H.~J. Warringa,
\newblock Nucl. Phys. A {\bf 803}, 227 (2008), arXiv:0711.0950.

\bibitem{Kharzeev:2007tn}
D.~Kharzeev and A.~Zhitnitsky,
\newblock Nucl. Phys. A {\bf 797}, 67 (2007), arXiv:0706.1026.

\bibitem{Fukushima:2008xe}
K.~Fukushima, D.~E. Kharzeev, and H.~J. Warringa,
\newblock Phys. Rev. D {\bf 78}, 074033 (2008), arXiv:0808.3382.

\bibitem{Fukushima:2010vw}
K.~Fukushima, D.~E. Kharzeev, and H.~J. Warringa,
\newblock Phys. Rev. Lett. {\bf 104}, 212001 (2010), arXiv:1002.2495.

\bibitem{Kharzeev2011}
D.~E. Kharzeev and D.~T. Son,
\newblock Phys. Rev. Lett. {\bf 106}, 062301 (2011), arXiv:1010.0038.

\bibitem{Son:2009tf}
D.~T. Son and P.~Surowka,
\newblock Phys. Rev. Lett. {\bf 103}, 191601 (2009), arXiv:0906.5044.

\bibitem{Son:2004tq}
D.~T. Son and A.~R. Zhitnitsky,
\newblock Phys. Rev. D {\bf 70}, 074018 (2004), arXiv:hep-ph/0405216.

\bibitem{Metlitski:2005pr}
M.~A. Metlitski and A.~R. Zhitnitsky,
\newblock Phys. Rev. D {\bf 72}, 045011 (2005), arXiv:hep-ph/0505072.

\bibitem{Kharzeev:2010gd}
D.~E. Kharzeev and H.-U. Yee,
\newblock Phys. Rev. D {\bf 83}, 085007 (2011), arXiv:1012.6026.

\bibitem{Burnier:2011bf}
Y.~Burnier, D.~E. Kharzeev, J.~Liao, and H.-U. Yee,
\newblock Phys. Rev. Lett. {\bf 107}, 052303 (2011), arXiv:1103.1307.

\bibitem{Burnier:2012ae}
Y.~Burnier, D.~E. Kharzeev, J.~Liao, and H.~U. Yee,
\newblock (2012), arXiv:1208.2537.

\bibitem{Yee:2013cya}
H.-U. Yee and Y.~Yin,
\newblock Phys. Rev. C {\bf 89}, 044909 (2014), arXiv:1311.2574.

\bibitem{Jiang2015}
Y.~Jiang, X.-G. Huang, and J.~Liao,
\newblock Phys. Rev. D {\bf 92}, 071501 (2015), arXiv:1504.03201.

\bibitem{Abelev:2009uh}
STAR, B.~I. Abelev {\em et~al.},
\newblock Phys. Rev. Lett. {\bf 103}, 251601 (2009), arXiv:0909.1739.

\bibitem{Abelev:2009txa}
STAR, B.~I. Abelev {\em et~al.},
\newblock Phys. Rev. {\bf C81}, 054908 (2010), arXiv:0909.1717.

\bibitem{Abelev2013}
ALICE, B.~Abelev {\em et~al.},
\newblock Phys. Rev. Lett. {\bf 110}, 012301 (2013), arXiv:1207.0900.

\bibitem{Adamczyk2013}
STAR, L.~Adamczyk {\em et~al.},
\newblock Phys. Rev. C {\bf 88}, 064911 (2013), arXiv:1302.3802.

\bibitem{Adamczyk2014}
STAR, L.~Adamczyk {\em et~al.},
\newblock Phys. Rev. Lett. {\bf 113}, 052302 (2014), arXiv:1404.1433.

\bibitem{Adamczyk2014a}
STAR, L.~Adamczyk {\em et~al.},
\newblock Phys. Rev. C {\bf 89}, 044908 (2014), arXiv:1303.0901.

\bibitem{Khachatryan2017}
CMS, V.~Khachatryan {\em et~al.},
\newblock Phys. Rev. Lett. {\bf 118}, 122301 (2017), arXiv:1610.00263.

\bibitem{Schlichting2011}
S.~Schlichting and S.~Pratt,
\newblock Phys. Rev. C {\bf 83}, 014913 (2011), arXiv:1009.4283.

\bibitem{Wang2017}
F.~Wang and J.~Zhao,
\newblock Phys. Rev. C {\bf 95}, 051901 (2017), arXiv:1608.06610.

\bibitem{Zhao2019}
STAR, J.~Zhao,
\newblock Nucl. Phys. A {\bf 982}, 535 (2019), arXiv:1807.09925.

\bibitem{Bzdak2013}
A.~Bzdak, V.~Koch, and J.~Liao,
\newblock Lect. Notes Phys. {\bf 871}, 503 (2013), arXiv:1207.7327.

\bibitem{Zhao:2017nfq}
J.~Zhao, H.~Li, and F.~Wang,
\newblock Eur. Phys. J. C {\bf 79}, 168 (2019), arXiv:1705.05410.

\bibitem{Xu2018}
H.-j. Xu {\em et~al.},
\newblock Chin. Phys. C {\bf 42}, 084103 (2018), arXiv:1710.07265.

\bibitem{Wang2010}
F.~Wang,
\newblock Phys. Rev. C {\bf 81}, 064902 (2010), arXiv:0911.1482.

\bibitem{Bloczynski:2012en}
J.~Bloczynski, X.-G. Huang, X.~Zhang, and J.~Liao,
\newblock Phys. Lett. B {\bf 718}, 1529 (2013), arXiv:1209.6594.

\bibitem{Bloczynski2015}
J.~Bloczynski, X.-G. Huang, X.~Zhang, and J.~Liao,
\newblock Nucl. Phys. A {\bf 939}, 85 (2015), arXiv:1311.5451.

\bibitem{Chatterjee2015}
S.~Chatterjee and P.~Tribedy,
\newblock Phys. Rev. C {\bf 92}, 011902 (2015), arXiv:1412.5103.

\bibitem{Voloshin2010}
S.~A. Voloshin,
\newblock Phys. Rev. Lett. {\bf 105}, 172301 (2010), arXiv:1006.1020.

\bibitem{STAR:2021mii}
STAR, M.~Abdallah {\em et~al.},
\newblock Phys. Rev. C {\bf 105}, 014901 (2022), arXiv:2109.00131.

\bibitem{Hu2022}
STAR, Y.~Hu,
\newblock EPJ Web Conf. {\bf 259}, 13013 (2022), arXiv:2110.15937.

\bibitem{Hu2023}
STAR, Y.~Hu,
\newblock Acta Phys. Polon. Supp. {\bf 16}, 44 (2023), arXiv:2208.09069.

\bibitem{Deng2016}
W.-T. Deng, X.-G. Huang, G.-L. Ma, and G.~Wang,
\newblock Phys. Rev. C {\bf 94}, 041901 (2016), arXiv:1607.04697.

\bibitem{Deng2018}
W. T. Deng, X.G. Huang, G. L. Ma, and G. Wang
\newblock Phys. Rev. {\bf C97}, 044901 (2018), arXiv:1802.02292.

\bibitem{Shi2019}
S. Shi, H. Zhang, D. Hou, and J. Liao
\newblock Nucl. Phys. A {\bf 982}, 539-542 (2019), arXiv:1807.05604. 

\bibitem{Zhao2019a}
X.-L. Zhao, G.-L. Ma, and Y.-G. Ma,
\newblock Phys. Rev. C {\bf 99}, 034903 (2019), arXiv:1901.04151.

\bibitem{McLerran:2013hla}
L.~McLerran and V.~Skokov,
\newblock Nucl. Phys. A {\bf 929}, 184 (2014), arXiv:1305.0774.

\bibitem{Tuchin:2013apa}
K.~Tuchin,
\newblock Phys. Rev. C {\bf 88}, 024911 (2013), arXiv:1305.5806.

\bibitem{Tuchin:2014iua}
K.~Tuchin,
\newblock Phys. Rev. C {\bf 91}, 064902 (2015), arXiv:1411.1363.

\bibitem{Li:2016tel}
H.~Li, X.-l. Sheng, and Q.~Wang,
\newblock Phys. Rev. C {\bf 94}, 044903 (2016), arXiv:1602.02223.

\bibitem{Siddique:2021smf}
I.~Siddique, X.-L. Sheng, and Q.~Wang,
\newblock Phys. Rev. C {\bf 104}, 034907 (2021), arXiv:2106.00478.

\bibitem{Siddique2022}
I.~Siddique, S.~Cao, U.~Tabassam, M.~Saeed, and M.~Waqas,
\newblock Phys. Rev. C {\bf 105}, 054909 (2022), arXiv:2201.09634.

\bibitem{Loizides:2014vua}
C.~Loizides, J.~Nagle, and P.~Steinberg,
\newblock SoftwareX {\bf 1-2}, 13 (2015), arXiv:1408.2549.

\bibitem{Alver:2006wh}
PHOBOS Collaboration, B.~Alver {\em et~al.},
\newblock Phys.Rev.Lett. {\bf 98}, 242302 (2007), arXiv:nucl-ex/0610037.

\bibitem{Miller:2007ri}
M.~L. Miller, K.~Reygers, S.~J. Sanders, and P.~Steinberg,
\newblock Ann. Rev. Nucl. Part. Sci. {\bf 57}, 205 (2007),
  arXiv:nucl-ex/0701025.

\bibitem{Alver:2008zza}
B.~Alver {\em et~al.},
\newblock Phys. Rev. {\bf C77}, 014906 (2008), arXiv:0711.3724.

\bibitem{Moller1995}
P.~Moller, J.~R. Nix, W.~D. Myers, and W.~J. Swiatecki,
\newblock Atom. Data Nucl. Data Tabl. {\bf 59}, 185 (1995),
  arXiv:nucl-th/9308022.

\bibitem{Pritychenko2016}
B.~Pritychenko, M.~Birch, B.~Singh, and M.~Horoi,
\newblock Atom. Data Nucl. Data Tabl. {\bf 107}, 1 (2016), arXiv:1312.5975,
\newblock [Erratum: Atom.Data Nucl.Data Tabl. 114, 371--374 (2017)].

\bibitem{Shou2015}
Q.~Y. Shou {\em et~al.},
\newblock Phys. Lett. B {\bf 749}, 215 (2015), arXiv:1409.8375.

\bibitem{Xu2021}
H.-j. Xu, H.~Li, X.~Wang, C.~Shen, and F.~Wang,
\newblock Phys. Lett. B {\bf 819}, 136453 (2021), arXiv:2103.05595.

\bibitem{Zhao2022}
X.-L. Zhao and G.-L. Ma,
\newblock Phys. Rev. C {\bf 106}, 034909 (2022), arXiv:2203.15214.

\bibitem{Qin:2010pf}
G.-Y. Qin, H.~Petersen, S.~A. Bass, and B.~Muller,
\newblock Phys. Rev. {\bf C82}, 064903 (2010), arXiv:1009.1847.

\bibitem{Teaney:2010vd}
D.~Teaney and L.~Yan,
\newblock Phys. Rev. {\bf C83}, 064904 (2011), arXiv:1010.1876.

\bibitem{Hattori2017}
K. Hattori, and X. G. Huang
\newblock Nucl. Sci. Tech. {\bf 28}, 26 (2017), arXiv:1609.00747.

\bibitem{Roy2015}
V. Roy, S. Pu, L. Rezzolla,and D. Rischke
\newblock Phys. Lett. B. {\bf 750}, 45-52 (2015), arXiv:1506.06620.

\bibitem{Pu2016}
S. Pu, V. Roy, L. Rezzolla, and D. Rischke
\newblock Phys. Rev. D. {\bf 93}, 074022 (2016), arXiv:1602.04953.

\bibitem{Yan2023}
L. Yan, and X. G. Huang
\newblock Phys. Rev. D. {\bf 107}, 094028 (2023), arXiv:2104.00831

\end{thebibliography}

\end{document}